\documentclass[useAMS,usenatbib]{mn2e}
\usepackage{graphicx}
\usepackage{amsmath}
\usepackage{amssymb}
\usepackage[greek,english]{betababel}

\def\be{\begin{equation}}
\def\ee{\end{equation}}
\def\bea{\begin{eqnarray}}
\def\eea{\end{eqnarray}}

\title[On the dissolution of star clusters in the Galactic centre. I. Circular orbits.]{On the dissolution of star clusters in the Galactic centre. I. Circular orbits.}
\author[A. Ernst, A. Just, R. Spurzem]{A. Ernst$^{1,2}$\thanks{email: aernst@ari.uni-heidelberg.de}, A. Just$^{1}$\thanks{email: just@ari.uni-heidelberg.de},
R. Spurzem$^{1}$\thanks{email: spurzem@ari.uni-heidelberg.de} \\
$^{1}$Astronomisches Rechen-Institut/Zentrum f\"ur Astronomie der Universit\"at Heidelberg, 
M\"onchhofstrasse 12-14,
69120 Heidelberg, Germany\\
$^{2}$Max-Planck-Institut f\"ur Astronomie, K\"onigstuhl 17, 69117 Heidelberg, Germany}

\begin{document}

\date{Accepted ... Received ...}

\pagerange{\pageref{firstpage}--\pageref{lastpage}} \pubyear{2002}

\maketitle

\label{firstpage}

\begin{abstract}
We present $N$-body simulations of dissolving star clusters close to galactic centres.
For this purpose, we developed a new $N$-body program called {\sc nbody6gc}
based on Aarseth's series of $N$-body codes. We describe the algorithm in detail.
We report about the density wave phenomenon in the tidal arms which has been recently 
explained by K\"upper et al. (2008). Standing waves develop in the tidal arms.  The wave knots 
or clumps develop at the position, where the emerging tidal arm hits the potential wall of the 
effective potential and is reflected. The escaping stars move through 
the wave knots further into the tidal arms. We show the consistency of the positions of the wave 
knots with the theory in Just et al. (2009). 
We also demonstrate a simple method to study the properties of tidal arms. By 
solving many eigenvalue problems along the tidal arms, we construct numerically a 1D coordinate system whose direction is always along a principal axis of the local tensor of inertia. Along this coordinate system, physical quantities can be evaluated. The half-mass or dissolution times of 
our models are almost independent of the particle number which indicates that two-body
relaxation is not the dominant mechanism leading to the dissolution. This may be a typical 
situation for many young star clusters. We propose a classification scheme which sheds
light on the dissolution mechanism.
\end{abstract}


\raggedbottom

\section{Introduction}

The centres of the Milky Way and other galaxies are currently a field of very intensive 
research.\footnote{As usual, we denote our Galaxy with a capital letter ``G'' while galaxies 
in general will be denoted with a lower-case ``g''.} 
In our Galaxy, observations have to be carried out in other wavelengths than visual due to the 
huge extinction. Directly in the centre of our Galaxy resides the strong radio source Sgr A* 
 at the location of the Galactic super-massive black hole ($M_\bullet\approx (3 - 4)\times 
 10^6\ M_\odot$, e.g. Genzel et al. 2000, Ghez et al. 2000, Sch\"odel et al. 2002, Ghez et al. 2003, 
 Eckart et al. 2005, Ghez et al. 2005, Beloborodov et al. 2006).
The Galactic centre region spans roughly {\it nine} orders of magnitude in galactocentric 
radii ranging from a rough outer radius of the central molecular zone 
($R_{\rm CMZ}\approx 200$ pc, Morris \& Serabyn 1996) 
down to the Schwarzschild radius of the Galactic super-massive black hole 
($R_{\bullet}\approx 4\times10^{-7}$ pc).
This large range in radial scales already suggests that the physics in the Galactic centre region is extremely rich in content.

Two young star-burst clusters named Quintuplet (Nagata et al. 1990, 
Okuda et al. 1990) and Arches (Nagata et al. 1995) have been discovered at projected 
distances less than $35$ pc away from the Galactic centre. They have quite 
extraordinary properties and stellar contents.
Their formation still requires clarification. However, both clusters are located (at least, in 
projection) near the Galactic centre ``Radio Arc''  (Yusef-Zadeh, Morris \& Chance 1984, 
Timmermann et al. 1996), which is a 
region rich in molecular clouds and gaseous filaments (Morris \& Serabyn 1996, Lang et al. 2005).

The tidal field is extremely strong in the Galactic centre. It was therefore highly desirable for 
us to study the effect of the tidal field on the dynamics of star clusters which
orbit around galactic centres at small galactocentric distances. 
A few similar simulations as those presented in this paper can be found in the works by 
Fujii et al. (2007, 2009). Other previous works on the dissolution of star clusters
in the Galactic centre have been published by Portegies Zwart, 
McMillan \& Gerhard (2003), Kim \& Morris (2003) and Guerkan \& Rasio (2005).

We do not attempt to solve the paradox of youth (Ghez et al. 2003) 
in this study with the star cluster in-spiral scenario (Gerhard 2001). 
Gerhard used $\ln\Lambda=10$ for the Coulomb logarithm of dynamical friction. 
This value is, from our point of view, much too large. 
We will use in this study a more realistic and variable Coulomb logarithm 
according to Just \& Pe\~narrubia (2005) which leads to in-spiral time scales 
which are considerably larger.

This paper is organised as follows: In Section 2, we describe in detail the algorithm
of our $N$-body program {\sc nbody6gc} which has been especially developed to study the
dynamics of star clusters in galactic centres. Section 3 describes the theoretical models which 
we use for the central region of the Galactic bulge and the star clusters. In addition, we discuss 
the effective potential which is essential in order to understand the dynamics in the tidal field, and
Poincar\'e surfaces of section. Section 4 contains the results of our direct $N$-body
simulations. Our main focus is on the properties of the tidal arms and the dissolution time. 
In Appendix \ref{sec:taylorexpansion}, we show the Taylor expansion of the effective potential.
In Appendix \ref{sec:tidalarmcoordinatesystem}, we describe the algorithm of an eigensolver which is used to construct a 1D 
coordinate system along the tidal arms. It can be used to evaluate physical quantities along the
tidal arms in order to study their properties. 

\section{Numerical method}

\begin{figure}
\centering
\includegraphics[width=0.5\textwidth]{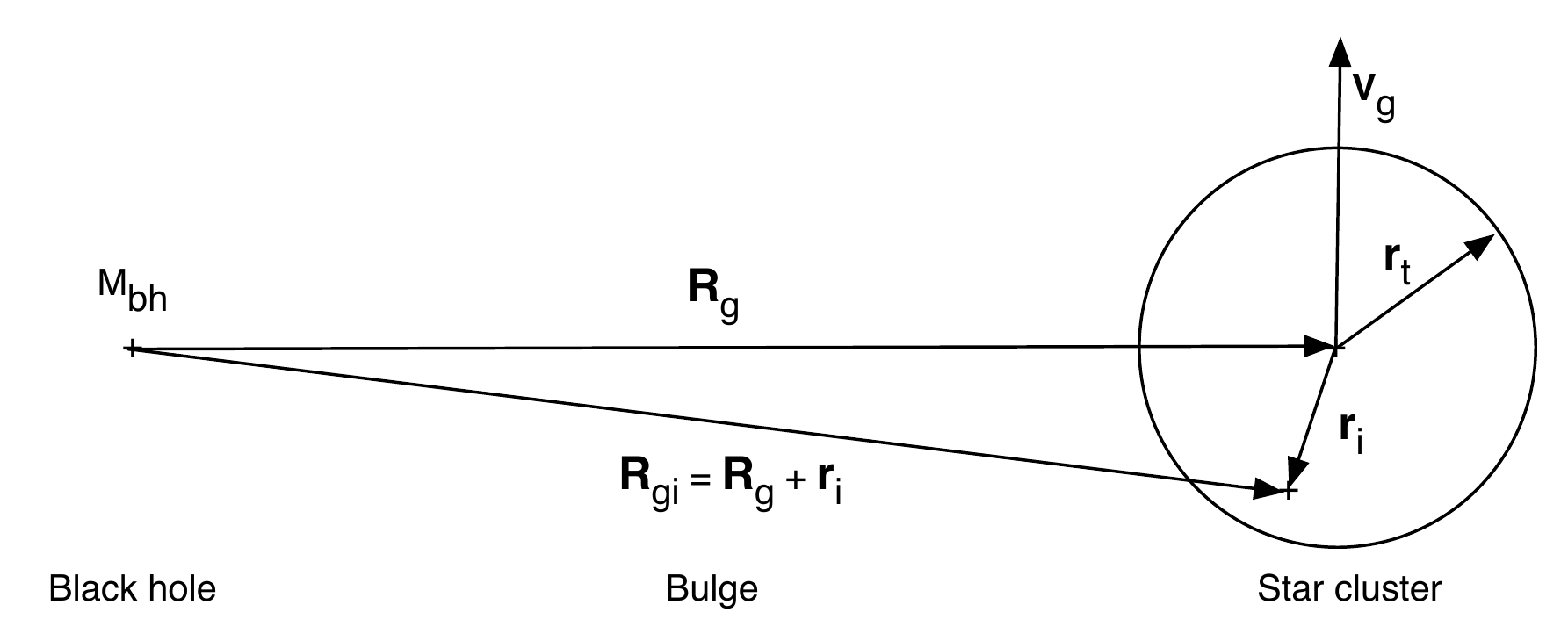} 
\caption{Sketch of the geometry of the problem as seen from the galactocentric reference frame. 
The vector 
$\mathbf{r}_g$ points from the Galactic centre to the star cluster centre. The vector $\mathbf{r}_i$
points from the star cluster centre to the position of the $i$th star. The orbital velocity of
the star cluster has been denoted as $\mathbf{V}_g$.} 
\label{fig:gcfig}
\end{figure}

The computer program {\sc nbody6gc} which is used in this study is a variant of the $N$-body program 
{\sc nbody6++} (Aarseth 1999, 2003, Spurzem 1999) suited for massively
parallel computers.\footnote{We remark here that {\sc nbody6gc} is based on a code 
variant called {\sc nbody6tid} which has been developed by 
R. Spurzem in collaboration with O. Gerhard and K.-S. Oh (unpublished). 
{\sc nbody6tid} was very helpful for the development of {\sc nbody6gc}. However, we switched 
to another integrator for circular and very eccentric cluster orbits and improved the treatment of dynamical friction for studies in the Galactic centre.} 
The code {\sc nbody6++}  is a variant of the direct $N$-body code {\sc nbody6} (Aarseth 1999, 2003)
for single-processor machines. A fourth-order Hermite scheme, applied first by Makino \& Aarseth (1992), is used  for the direct integration of the $3N$ Newtonian equations of motion of the $N$-body 
system. It uses adaptive and individual time steps, which are organised in hierarchical 
block time steps, the Ahmad-Cohen neighbour scheme (Ahmad \& Cohen 1973),
Kustaanheimo-Stiefel (KS) regularisation of close encounters
(Kustaanheimo \& Stiefel 1965) and Chain regularisation (Mikkola \& Aarseth 1990, 1993, 1996,1998). 

\subsection{Cluster orbit}

We denote the radii, velocities and accelerations related to the Galactic centre with capital letters 
and those related to the star cluster with lower case letters. Figure \ref{fig:gcfig} shows the geometry of the problem: A star cluster is orbiting around the Galactic centre. The potential in which the star cluster moves is the sum of the Kepler potential of a super-massive black hole and a scale free potential of the central 
region of the Galactic bulge (cf. Section \ref{sec:scalefree}).\footnote{The program 
{\sc nbody6gc} is written in a way that any analytical galactic potential can be 
implemented.}  
The two first-order equations of motion 
for the star cluster orbit read

\bea
\mathbf{V}_g(t) &=& \dot{\mathbf{R}}_g , \label{eq:1steqm} \\
\dot{\mathbf{V}}_g(t) &=& -\nabla\Phi_{g}(\vert \mathbf{R}_g\vert) + \mathbf{A}_{\rm df}, \label{eq:2ndeqm}
\eea

\noindent
where $\mathbf{R}_g, \mathbf{V}_g, \Phi_{g}$ and $\mathbf{a}_{\rm df}$ are the position vector, velocity vector, gravitational potential of the Galactic centre region and deceleration due to dynamical friction
and the dot denotes the derivative with respect to time. 
The equations of motion 
(\ref{eq:1steqm}) and (\ref{eq:2ndeqm}) for the star cluster orbit with respect to the Galactic centre are solved using an
 $8$th-order composition scheme (McLachlan 1995; for the idea see Yoshida 1990) 
 with implicit midpoint method (e.g. Mikkola \& Aarseth 2002), thereby including a realistic dynamical friction force. Although the symplectic composition schemes are by construction suited for
Hamiltonian systems, they can be used for dissipative systems as well if the dissipative force is 
not too large. In our case, four iterations turned out to be sufficient to guarantee an excellent accuracy
of the scheme.

 We use a cluster membership criterion such that the dynamical friction force is based 
 only on the total mass of the cluster members. We define a membership radius $r_m$ 
 by the condition

\be
\overline{\rho}_{cl} = \frac{3M_{cl}(r_m)}{4\pi r_m^3} = \rho_g(R_g) \label{eq:rtdens}
\ee

\noindent
as the radius where the mean density $\overline{\rho}_{cl}$ in the star cluster is equal to the
local bulge density at the cluster centre which is located at radius $R_g$. This radius differs
from the tidal radius (King 1962) only by a factor of order unity. Stars within
twice the membership radius are defined as cluster members.

\subsection{Stellar orbits}

On the other hand, the equations of motion for the orbits of stars in the star cluster are solved by
the standard {\sc nbody6/nbody6++} routines using the
$4$th-order Hermite scheme (Makino \& Aarseth 1992), KS or chain regularisation 
including the full 3D tidal forces from the super-massive black hole and the Galactic bulge. 
The tidal force is added as a perturbation to the KS regularisation. 
The following quantities are involved:
 
 \begin{enumerate}

\item The specific force on the $i$th particle due to all other stars (cluster members and 
non-members) is given by

\be
\mathbf{a}_i = G \sum_{\substack{j=1\\ j \not= i}}^N m_j
 \frac{\mathbf{r}_{ji}}{\vert \mathbf{r}_{ji} \vert^3}, \label{eq:ai}
\ee

where $N$, $G$, $\mathbf{r}_{ji} = \mathbf{r}_j - \mathbf{r}_i$, $m_j$ are the particle number, the gravitational constant, the relative position vector between the $i$th and $j$th particles and the 
mass of the $j$th particle, respectively.

\item The specific force due to the Galactic centre at the position of the cluster centre is
 
 \be
 \mathbf{A}_{g} =- \left( \frac{GC}{R_{g}^{2-\alpha}} + \frac{GM_{bh}}{R_{g}^2} \right)  \frac{\mathbf{R}_{g}}{R_{g}},
 \ee
 
 \noindent
 where $C$, $M_{bh}$ and $\alpha$ are the normalisation of the scale free bulge mass profile 
 (see Section \ref{sec:scalefree}), the
mass of the super-massive black hole and the cumulative mass profile power law index. 

\item The specific force exerted on particle $i$ due to the Galactic centre is given by
 
\be
\mathbf{A}_{gi} = - \left( \frac{GC}{R_{gi}^{2-\alpha}} + \frac{GM_{bh}}{R_{gi}^2} \right)  \frac{\mathbf{R}_{gi}}{R_{gi}},
\ee

\item The deceleration due to dynamical friction is given by

\be
\mathbf{A}_{df} = -\frac{4\pi G^2\rho_g M_{cl}}{V_g^2} \ln\Lambda \, \chi(V_g) \, \frac{\mathbf{V}_g}{V_g}\label{eq:adf}
\ee

\noindent
where $\rho_g$, $M_{cl}$, $\mathbf{V}_g$ and $V_g$ are the local bulge density at the position of the star cluster centre, the star cluster mass and the velocity vector and modulus of the Galactic centre,
respectively.
Furthermore, $\ln\Lambda$ is the Coulomb logarithm which results from the integral over impact parameters and $\chi(v_g)=\int_0^{V_g} f(v) d^3v$ is the result of the integration of the distribution function $f(v)$ of light particles over velocity space. For the Coulomb logarithm $\ln\Lambda$, we use
according to Just \& Pe\~narrubia (2005)

\bea
\ln\Lambda &=& \ln\left(\frac{b_{1}}{b_{0}}\right), \label{eq:lnl} \\ 
b_1^2 &=& b_0^2 + L^2, \ \ \ \ \ b_0 = r_V, \ \ \ \ \ L = \frac{\rho_g}{\nabla\rho_g}  \label{eq:lnl2}
\eea

\noindent
where $b_1$, $b_0$, $L$ are the maximum and minimum impact parameters and
the local scale length of the bulge density profile, respectively, and  
$r_V = GM_{cl}^2/(4\vert E_{cl}\vert) \approx r_h$ is the virial 
radius of the star cluster (where $E_{cl}$ is the internal energy of the star cluster and $r_h$ is
the half-mass radius).
\end{enumerate}

In the galactocentric reference frame, the total force on the $i$th particle would be given by

\bea
\mathbf{A}_{tot,i,gc} &=& \mathbf{a}_{i,gc} + \mathbf{A}_{gi,gc} + \mathbf{A}_{df,gc}  \ \ \ \ \ \mathrm{memb.} \\
\mathbf{A}_{tot,i,gc} &=&  \mathbf{a}_{i,gc} + \mathbf{A}_{gi,gc} \ \ \ \ \ \ \ \ \ \ \ \ \ \ \ \ \, \mathrm{non-memb.}
\eea

\noindent
where the subscript ``gc'' denotes ``galactocentric''.
However, we choose the cluster rest frame as reference frame for our simulations. This is necessary,
because Aarseth's family of $N$-body programs is adapted to this reference frame and
assumes that the cluster centre is close to the origin of coordinates. This guarantees a 
sufficient accuracy of the Hermite scheme which is used for the orbit integration.
On the other hand, this choice of the reference frame implies that the Galactic centre is 
modelled as a pseudo-particle which orbits around the cluster centre.
We keep in mind that a transformation from the galactocentric frame to the cluster rest frame implies 
that $\mathbf{r}_i$, $\mathbf{R}_g$, $\mathbf{V}_g$ and $\mathbf{R}_{gi}$ in 
(\ref{eq:ai}) - (\ref{eq:adf}) change their sign. This implies that

\bea
\mathbf{a}_{i,cl} &=& -\mathbf{a}_{i,gc}, \ \ \ \ \   \mathbf{A}_{g,cl} = - \mathbf{A}_{g,gc}, \\
\mathbf{A}_{gi,cl} &=& - \mathbf{A}_{gi,gc},
\ \ \ \ \ \mathbf{A}_{df,cl} = -\mathbf{A}_{df,gc}
\eea

\noindent
where the subscript ``cl'' denotes the cluster frame.
It is then convenient for the force computations to transform to a reference frame
in which the initial cluster centre is force-free. Since this frame is accelerated,
an apparent force

\bea
\mathbf{A}_{app} &=& - \mathbf{A}_{g,cl} - \mathbf{A}_{df,cl} \ \ \ \ \  \mathrm{memb.} \\
\mathbf{A}_{app} &=& - \mathbf{A}_{g,cl} \ \ \ \ \ \ \ \ \ \ \ \ \ \ \ \ \mathrm{non-memb.}
\eea

\noindent
appears. In the accelerated cluster frame the total force on the $i$th particle is therefore given by

\be
\mathbf{A}_{tot,i,acl} = -\mathbf{A}_{tot,i,gc} + \mathbf{A}_{app} 
\ee

\noindent
where the subscript ``acl'' denotes the accelerated cluster frame.
Thus the second-order equations of motion for the orbits of the cluster stars read

\bea
\mathbf{a}_{tot,i,acl} &=& \mathbf{a}_{i,cl} + \mathbf{A}_{gi,cl} - \mathbf{A}_{g,cl} \ \ \ \ \ \ \ \ \ \ \ \ \ \ \  \mathrm{memb.} \\
\mathbf{a}_{tot,i,acl} &=&  \mathbf{a}_{i,cl} + \mathbf{A}_{gi,cl} -\mathbf{A}_{g,cl} - \mathbf{A}_{df,cl}  \ \ \ \,   \mathrm{n.-m.}
\eea

\noindent
It can be seen that in the accelerated cluster frame an individual star experiences only the 
differential tidal force between its own location and the cluster centre.

We applied a density centre correction in certain intervals to correct for the displacement
of the density centre. This was done in order to retain a consistent 
treatment of dynamical friction since the dynamical friction force is determined from
the approximation that the star cluster mass is concentrated in the origin of
coordinates.

\subsection{Energy check}

\label{sec:energycheck}

The specific energy $e_i$ of a particle is calculated in the galactocentric reference frame:

\be
e_i = \frac{1}{2} (\mathbf{v}_i + \mathbf{V}_g)^2 + \Phi_{i,int} + \Phi_{i,ext} + \int_{t_0}^{t_1} \mathbf{a}_{df}\cdot \mathbf{V}_{gi} \, dt \label{eq:specenergy}
\ee

\noindent
where $\Phi_{i,int} = -G\sum_{j=1 (\not= i)}^N m_j/{(\vert \mathbf{r}_i-\mathbf{r}_j}\vert )$
is the full internal potential of the $N$-body system, $\Phi_{i,ext}$ is the external potential
of the Galactic centre and the last term is the energy loss due to dynamical friction.
Using these terms, we calculate a total energy $E_1$. A factor $1/2$ has to be included
in the summation of the potential energy of the $N$-body system. Note that the individual terms
in (\ref{eq:specenergy}) have quite different orders of magnitude.
On the other hand, the orbital energy $E_2$ of the star cluster in the galactocentric reference 
frame is given by

\bea
E_2 &=& \frac{1}{2} M_{cl} \mathbf{V}_g^{\, 2} + M_{cl} \Phi_{ext}\left(\vert\mathbf{R}_g\right\vert) \nonumber \\
&&+ \int_{0}^{t} M_{cl} \, \mathbf{a}_{df} \cdot \mathbf{V}_g \, dt  + \int_{0}^{t} \frac{\dot{M_{cl}}}{M_{cl}} E_2 \, dt
\eea

\noindent
where the last two terms on the right-hand side are corrections due to dynamical friction and tidal 
mass loss of the star cluster, respectively. Both energies $E_1$ and $E_2$ are checked at regular 
intervals for conservation. 
We note that the problem of a star cluster orbiting in a galactic tidal field has two energy 
scales related to the internal energy of the star cluster and the external energy of the tidal field. 
Tidal heating can transfer external energy from the larger scale into internal energy. 
The program {\sc nbody6gc} conserves the energy within the cluster to 
a sufficient degree such that two-body relaxation is not suppressed (cf. Ernst 2009
for more details).

\section{Theory}

\begin{figure*}
\includegraphics[width=0.9\textwidth]{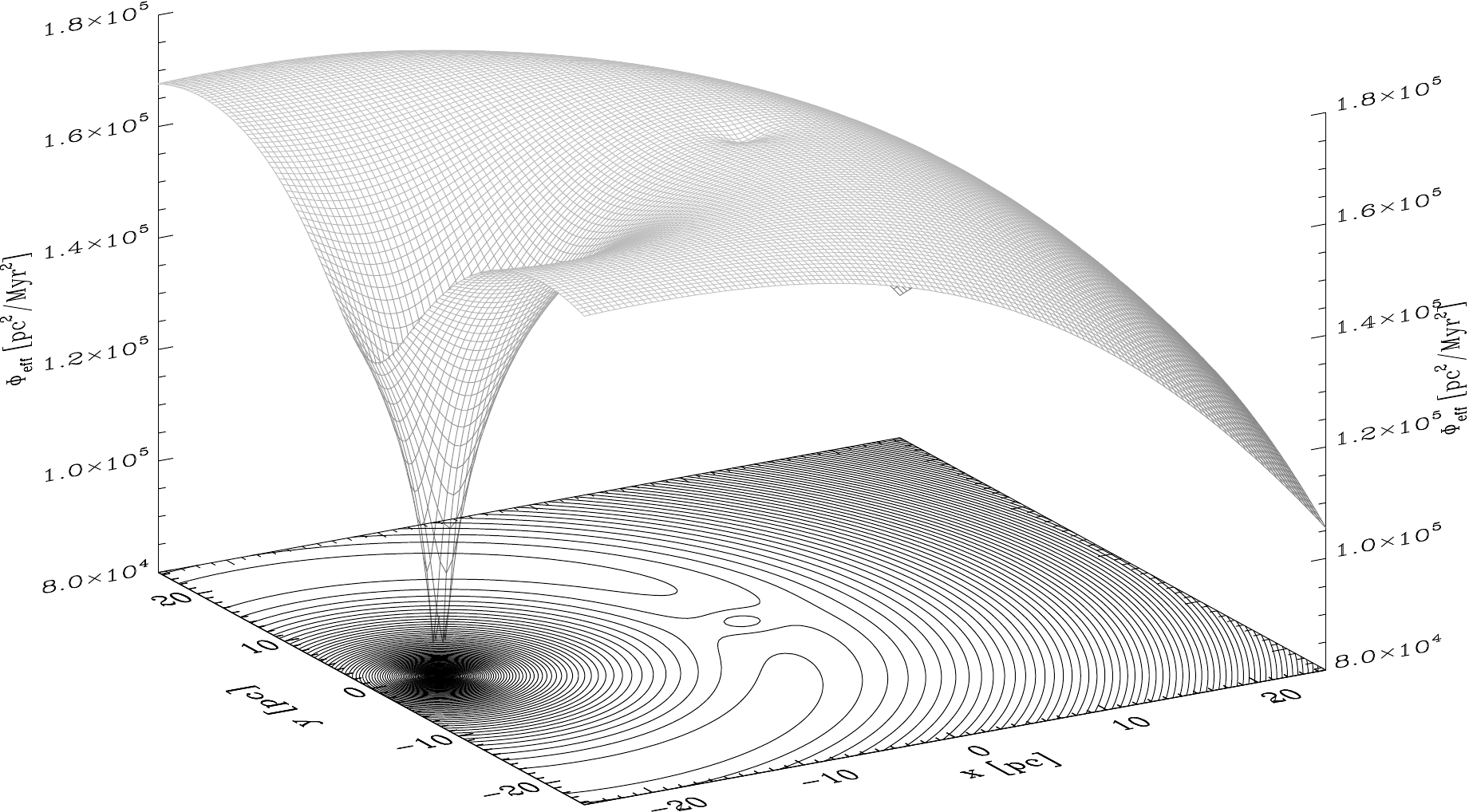} 
\caption{Effective potential for a star cluster in the centre of a galaxy ($z=0$ plane). The large 
potential well is due to the Galactic centre and the small one is due to the star cluster.} 
\label{fig:effpotgc}
\end{figure*}

\subsection{Galactic centre model}
\label{sec:scalefree}

In the following Sections, we will use parameter values close to those of the centre of the 
Milky Way, since these parameters are better known than those of any other centre of a galaxy.

For the very central region of the Galactic bulge, we use a spherically symmetric scale free 
model (i.e. a model which is self-similar under scaling of lengths). 
The potential $\Phi$, cumulative mass $M$ and density $\rho$ 
are given by

\bea
\Phi(u) &=& \left\{\begin{array}{l}
 \Phi_0 \, u^{\alpha - 1} \ \ \ \ \ \alpha \not=1, \label{eq:phiscalefree}\\
\Phi_0' \, \ln u \ \ \ \ \ \ \alpha = 1
\end{array} \right. \\
M(u) &=& M_0 \, u^\alpha  \label{eq:powermass}\\
\rho(u) &=& \rho_0 \, u^{\alpha-3} \label{eq:powerdensity}
\eea

\noindent
where 

\bea
u &=& R/R_0, \ \ \ \ \ \rho_0 = \frac{\alpha}{4\pi} \frac{M_0}{R_0^3}, \\
\Phi_0&=&\frac{1}{\alpha-1}\frac{GM_0}{R_0} = \frac{4\pi G}{\alpha (\alpha-1)} \rho_{0} R_0^2, \label{eq:phi0scalefree}
\eea

\noindent
$\alpha$ is the power law exponent of the cumulative mass profile, 
$G$ is the gravitational constant and $R_0$ is a length unit
(which is not inherent in nature but simply a human convention).



\noindent
The circular frequency $\omega$ is given by

\be
\omega(u) =  \omega_0 u^{(\alpha-3)/2}, \ \ \ \ \ \omega_0 = \sqrt{\frac{4\pi G \rho_0}{\alpha}} \label{eq:omegau}
\ee

\noindent
The ratio of the epicyclic frequency $\kappa$ to the circular frequency $\omega$ is given by

\be
\beta = \frac{\kappa}{\omega} = \sqrt{2\left[ \frac{d\ln\omega}{d\ln u} + 2\right] } = \sqrt{\alpha + 1} \label{eq:beta}
\ee

\noindent
The angular momentum of the circular orbit is given by

\be
L(u) = L_0 u^{(\alpha+1)/2}, \ \ \ \ \ L_0 = \omega_0 R_0^2.
\ee




%

\begin{table}
\begin{center}
\begin{tabular}{|l|l||l|l|}      
\hline
\hline
Parameter & Value & Parameter & Value\\
\hline
$\alpha$ & 1.2 & $E_J(L_1)$ [pc$^2$Myr$^{-2}$] & 1.67546e5 \\
$R_0$ [pc] & 20 & $E_J(L_2)$ [pc$^2$Myr$^{-2}$] & 1.67592e5 \\
$M_0$ [$M_\odot$] & 1.67459e8 & $E_J(L_3)$ [pc$^2$Myr$^{-2}$] & 1.65965e5 \\
$\rho_{0}$ [$M_\odot$pc$^{-3}$] & 1998.90 & $\Phi_{\rm eff,tid}(R_0)$ [pc$^2$Myr$^{-2}$] & 1.69502e5 \\
$\Phi_{0}$ [pc$^2$ Myr$^{-2}$] & 1.88335e5 & $x(L_1)$ [pc] & -2.66618 \\
$\omega_0$ [Myr$^{-1}$] & 9.704 & $x(L_2)$ [pc] & 2.78522 \\
$M_1$ [$M_\odot$] & $10^6$  & & \\
$r_1$ [pc] & $1.20213$ & $G$ [pc$^3$ $M_\odot^{-1}$ Myr$^{-2}$] &$(222.3)^{-1}$ \\
\hline
\end{tabular}
\end{center}
\caption{Parameters used for the model of the Galactic centre region 
(Section \ref{sec:scalefree}) and the Plummer models (Section \ref{sec:plummermodel})
which are used in Sections \ref{sec:effpotgc} and \ref{sec:poingc2}.
$C(L_i)$ and $x(L_i)$ the value of the effective potential at the Lagrange point $L_i$ 
and its location, respectively
and $G$ is the gravitational constant.}
\label{tab:paramanalytic}
\end{table}

The parameters of our models are given in Table \ref{tab:paramanalytic}.
The value of $M_0$ corresponds to $M_0(R_0=1 \ \mathrm{pc}) = 4.6 \times 10^6 M_\odot$.  
The stellar mass within the central parsec is not easy to determine (see
Sch\"odel et al. 2007, Genzel et al. 2003 and also the review by Mezger, Duschl \& Zylka 1996).

The simple numerical calculations in Sections \ref{sec:effpotgc} and 
\ref{sec:poingc2} have been done without a black hole at the Galactic centre.
Nevertheless, in our $N$-body calculations, we added the contribution of a super-massive black 
hole of mass $M_\bullet=3.6 \times 10^6 M_\odot$ (Eisenhauer et al. 2005).
However, the influence radius (Frank \& Rees 1976) of the Galactic super-massive black 
hole is only $1-2$ pc which is small compared to the galactocentric radii used 
in this study.  

\subsection{Star cluster model}

\label{sec:plummermodel}

For the star clusters, we use Plummer models for the simple numerical calculations
in Sections \ref{sec:effpotgc} and \ref{sec:poingc2}
and King models (King 1966) for all $N$-body models in Section \ref{sec:results}.

The Plummer model is given by 

\bea
\Phi(v) &=& -\Phi_1\frac{1}{\sqrt{1+v^2}}  \label{eq:plummerpotential} \\
M(v) &=& M_1\frac{v^3}{\left[1+v^2\right]^{3/2}}  \label{eq:plummermass} \\
\rho(v) &=& \rho_1 \frac{1}{\left[1+v^2\right]^{5/2}} \label{eq:plummerdensity}
\eea

\noindent
with the dimensionless radius $v=r/r_1\geq 0$ and the Plummer radius

\be
r_1 = \frac{GM_1}{\Phi_1} = \left( \frac{3M_1}{4\pi\rho_1}\right)^{1/3}
\ee

\noindent
where $M_1$ is the total cluster mass, $-\Phi_1$ is the central potential and $\rho_1$ the central 
density, all of them being finite. The parameters of the Plummer models are given
in Table \ref{tab:paramanalytic}.
The parameters of the King models which we use for the $N$-body simulations
are given in Table \ref{tab:clustermodels}.

 \begin{table}
\begin{center}
\begin{tabular}{l}
\hline
\hline
King models ($W_0 = 6$): \\
\hline
In general: $M_{\rm cl} = 10^6 M_\odot, r_h = 1.64 \ {\rm pc}, r_t = 11.2 \ {\rm pc}$; \\
K1 ($N=10^3$), K2 ($N=2\times 10^3$), K3 ($N=5\times 10^3$), \\
K4 ($N=10^4$), K5 ($N=2\times 10^4$), K6 ($N=3\times 10^4$), \\
K7 ($N=5\times 10^4$), K8 ($N=7\times 10^4$), K9 ($N=10^5$) \\
\hline
\end{tabular} 
\end{center}
\caption{Parameters of the $N$-body runs with King models. $W_0, c$ and $r_h$ are 
the dimensionless central potential, the concentration and the half-mass radius of the
King model, respectively, $N$ is the particle number and $M_{\rm cl}$ is the total cluster mass
(which is only needed to calculate a dynamical friction force).}
\label{tab:clustermodels}
\end{table}

\subsection{Effective potential}

\label{sec:effpotgc}

\begin{figure}
\includegraphics[angle=90,width=0.5\textwidth]{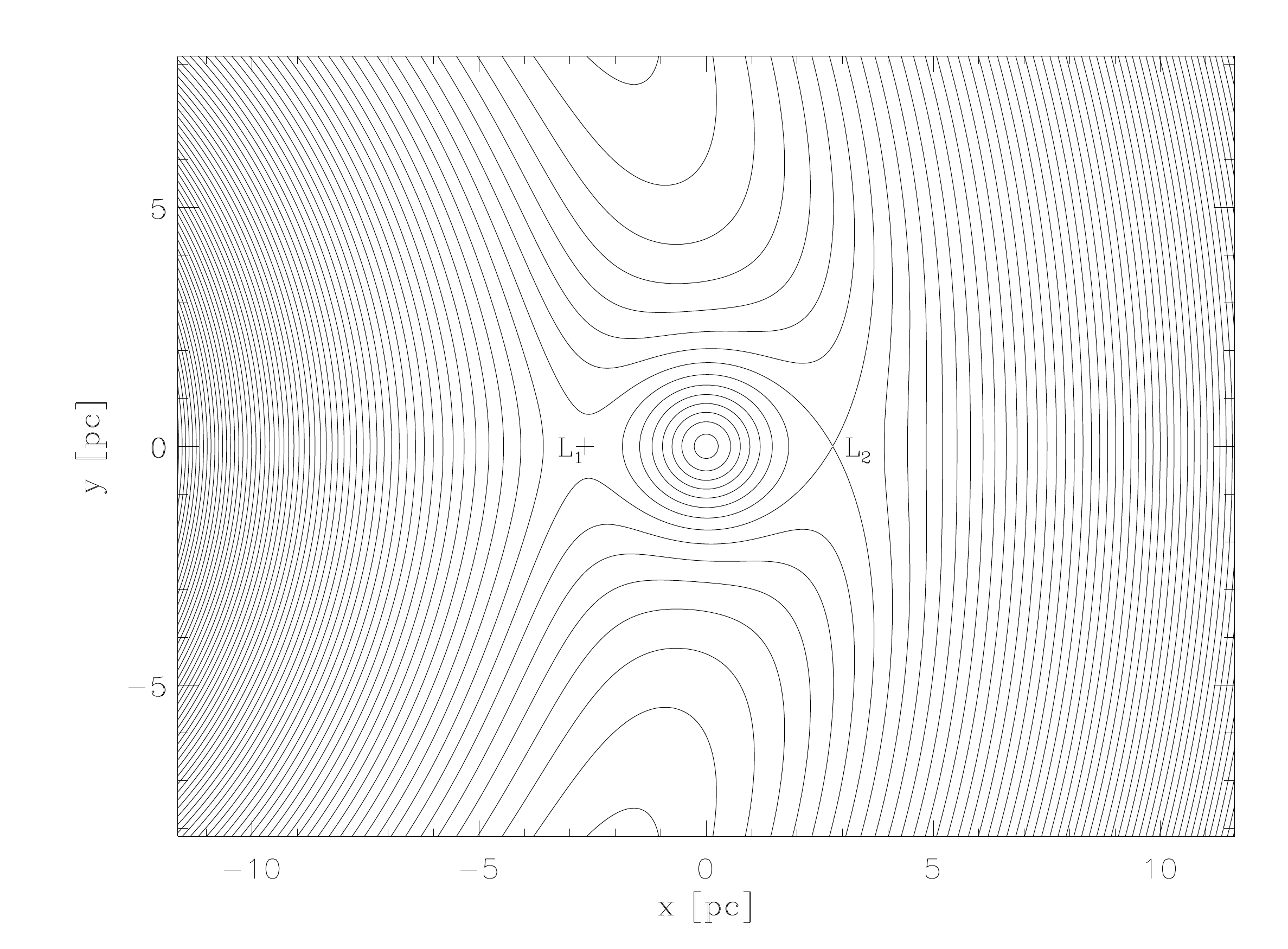} 
\caption{Zoom into the star cluster region of Figure \ref{fig:effpotgc}.} 
\label{fig:effpot3}
\end{figure}

\begin{figure}
\centering
\includegraphics[width=0.5\textwidth]{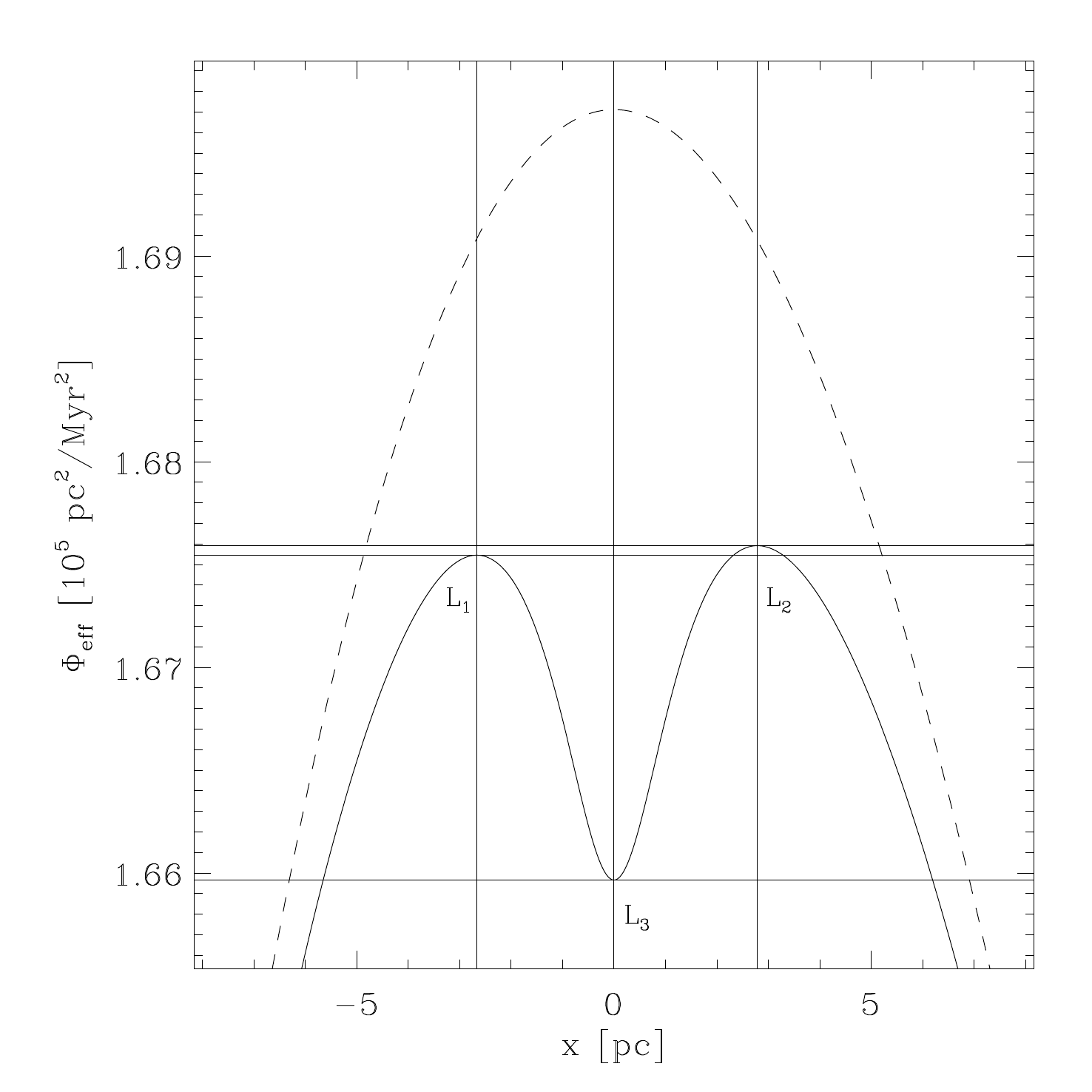} 
\caption{Zoom into the star cluster region of Figure \ref{fig:effpotgc} along the $x$ axis
with $y=0$. The Galactic centre lies in negative $x$ direction. It can be
seen that the Lagrangian points $L_1$ and $L_2$ lie at different energies and
at different distances from the cluster centre due to the asymmetry of the effective potential
with respect to $x=0$. The dashed line marks the effective tidal potential. The solid
line is the full potential with the added contribution of a Plummer potential.} 
\label{fig:gczoom2d}
\end{figure}

In this and the following Section, we use the parameters given in Table \ref{tab:paramanalytic}.
The length unit $R_0$ corresponds to the radius $R_C$ of the circular orbit, i.e. we have
$R_0=R_C$. 

Since we are considering a circular orbit, it is convenient to study the physics
in a reference frame which is co-rotating with the frequency $\omega_C=\omega_0$ 
of the circular orbit.
The star cluster centre is taken as the origin of coordinates. We choose a right-handed
coordinate system where the $x$-axis points away from the Galactic centre and the $y$-axis 
points in the orbital direction of the star cluster orbit around the Galactic centre. 
In this reference frame, centrifugal and Coriolis forces naturally appear according to 
classical mechanics. The potential in which a particle moves is the
superposition of the effective tidal potential and the star cluster potential. 
For short we will call this the effective potential.

The effective potential is shown in Figure \ref{fig:effpotgc}. It is given by the expression

\bea
\Phi_{\rm eff}\left(x,y,z\right) &=& \Phi_0 \left( \frac{\sqrt{(x+R_0)^2+y^2+z^2}}{R_0} \right)^{\alpha-1} \nonumber\\
&-& \frac{1}{2}\omega_0^2\left[ \left(x+R_0\right)^2 + y^2 + z^2 \right] \nonumber \\
&-& \frac{GM_1}{\sqrt{R_1^2+x^2+y^2 + z^2}} \label{eq:effpotgc}
\eea

\noindent
Note that $\Phi_0, \omega_0$ and $R_0$ are related by 
$\Phi_0 = \omega_0^2 R_0^2 / (\alpha - 1)$. 
In Equation (\ref{eq:effpotgc}), the first term is the gravitational
potential of the central region of the Galactic bulge, the second term is the centrifugal
potential and the last term is the Plummer potential of a star cluster. 
Note that the bulge potential is not well behaved in the limit $R_0 \rightarrow 0$ if there
 is no black hole.
However, the physics considered in this work happens close to the radii of the 
circular orbits. 
We note that the Jacobi energy per unit mass 
$E_J = \left( \dot{x}^2 + \dot{y}^2 + \dot{z}^2 \right)/2 + \Phi_{\rm eff}(x,y,z)$
is a conserved quantity in the co-rotating reference frame.

The tidal terms (i.e. the first two terms on the right-hand side) of Equation (\ref{eq:effpotgc}) can 
be expanded in a Taylor series around the star cluster centre $(x,y,z)=(0,0,0)$. Up to the $5$th order, the solution is given in Appendix \ref{sec:taylorexpansion}. The expansion up to the second order coincides with 
the tidal approximation which is a linear approximation of tidal forces. This approximation can be 
used to study the dynamics in star clusters on circular orbits which are far away from the Galactic 
centre. We stress, however, that we used the exact expressions for all computations in this
study.

Figure \ref{fig:effpot3} shows a zoom into the equipotential lines around the star cluster
region of Figure \ref{fig:effpotgc}. Figure \ref{fig:effpot3} also shows the location of the Lagrange points $L_1$ and $L_2$. As usual, $L_1$ lies on the negative $x$-axis (between the cluster centre and the Galactic centre) while $L_2$ lies on the positive $x$-axis. $L_1$ and $L_2$ 
are saddle points of the effective potential. It can be seen that at the locations of $L_1$ and $L_2$, the surface in Figure  \ref{fig:effpotgc} is curved differently along the $x-$ and $y-$axes. 

Figure \ref{fig:gczoom2d} shows the effective potential in the star cluster region along the 
line connecting the Galactic centre with the star cluster centre.  The dashed line shows only the effective
tidal potential $\Phi_{\rm eff,tid}$. The corresponding 1D Taylor series of $\Phi_{\rm eff,tid}$ along the 
$x$-axis around $x=0$ ($y=z=0$) is given by the power series

\bea
\Phi_{\rm eff,tid} &\approx& \frac{1}{2} \left(\frac{3-\alpha}{\alpha-1}\right)\omega_0^2 R_0^2 
+ \frac{1}{2}(\alpha - 3) \omega_0^2 x^2 \nonumber \\
&&+ \sum_{k=3}^\infty \left\{ \left[ \prod_{l=2}^k (\alpha - l)\right] \frac{\omega_0^2}{R_0^{k-2}} \frac{x^k}{k!}\right\}
\eea

\noindent
Higher-order terms lead to an asymmetry with respect to $x=0$ which becomes
important in the vicinity of the Galactic centre. A non-linearity in the tidal forces is
related to this asymmetry. Such non-linear effects can be seen in Poincar\'e surfaces of 
section. The solid line in Figure \ref{fig:gczoom2d} 
shows the full effective potential. It is the superposition of the effective tidal potential and the 
star cluster potential. The Lagrange points $L_1$ and $L_2$ lie at slightly different energies and 
at slightly different distances from the star cluster centre whose position we denoted as as $L_3$.
The energies and locations of the Lagrange points are given in Table \ref{tab:paramanalytic}.

We stress that this picture is only valid for a star cluster orbit which is exactly circular.
The region above the tidal effective potential (dashed line in Figure \ref{fig:gczoom2d}) 
is energetically forbidden for the cluster orbit. As soon as it becomes eccentric, the cluster centre 
no longer remains at the position of the extremum of the effective tidal potential but oscillates
around $x=0$ and is reflected either at the centrifugal or the gravitational barrier.
This oscillation leads to oscillations of the Jacobi energies of the Lagrangian points $L_1$
and $L_2$ on the orbital time scale of the star cluster orbit and can change the dynamics
dramatically.

\subsection{Poincar\'e surfaces of section}

\label{sec:poingc2}

\begin{figure*}
\centering
\includegraphics[width=0.9\textwidth]{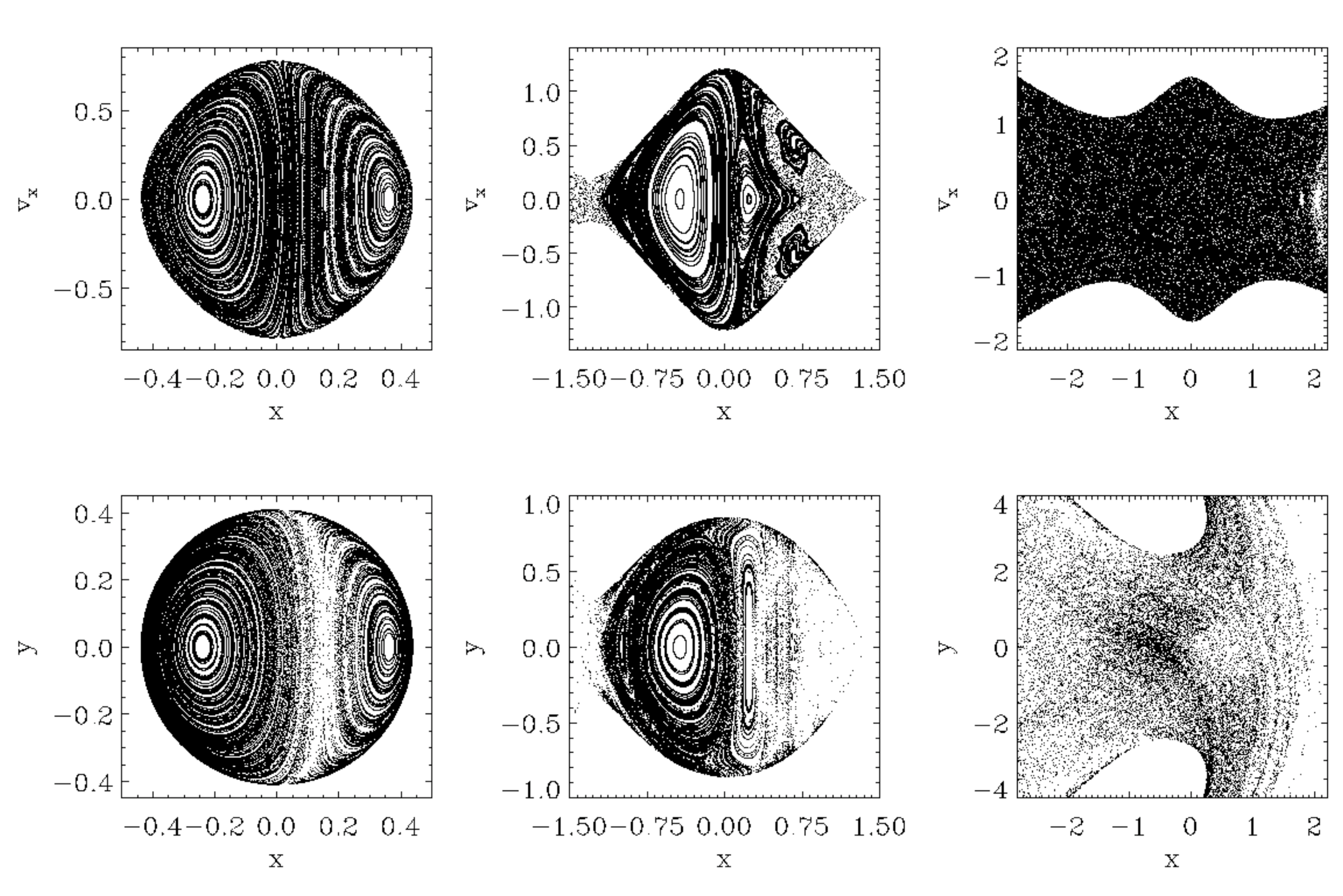} 
\caption{Poincar\'e surfaces of section. Left column: Deep in the potential well of the star cluster
at $C=1.66638e5$ pc$^2$/Myr$^2$. 
Middle column: At $E_J=E_J(L_2)$.
Right column: Above the Jacobi energies of $L_1$ and $L_2$ at $E_J=1.68845e5$ pc$^2$/Myr$^2$.}
\label{fig:poingc2}
\end{figure*}

Figure \ref{fig:poingc2} shows a few Poincar\'e surfaces of section for the orbit
with Parameters given in Table \ref{tab:paramanalytic} which is
exactly circular. 

The left column of Figure \ref{fig:poingc2} shows two Poincar\'e surfaces of section at 
a Jacobi energy deep in the potential well of the star cluster. 
The equipotential line corresponding to this Jacobi energy 
(which corresponds to the envelope of the lower surface of section in the 
left column of Figure \ref{fig:poingc2}) almost has a circular shape.
The Poincar\'e surfaces of section at this Jacobi energy show that all orbits are
regular and confined to invariant curves by a third integral. Such a third integral can 
usually be represented by a power series expansion where the lowest order is the
angular momentum which would be exactly conserved if the system were
spherical. Since the system is in fact not exactly spherical, the angular momentum
slightly oscillates around some value (see e.g. Figure 3-5 in Binney \& Tremaine 1987). 

The middle column of Figure \ref{fig:poingc2} shows two Poincar\'e surfaces of 
section at the Jacobi energy $E_J=E_J(L_2)$ which corresponds to the Lagrange
point $L_2$. The equipotential lines are open around $L_1$ and particles
can escape towards the Galactic centre. The phase space is divided between
regular and chaotic regions.

The right column of Figure \ref{fig:poingc2} shows two Poincar\'e surfaces of section 
at a Jacobi energy which is higher than the value of the effective potential at
both Lagrange points $L_1$ and $L_2$. The equipotential lines are wide open 
around $L_1$ and $L_2$ and particles can escape in both directions
either into the leading or the trailing tidal arm. All orbits are chaotic.

\section{Results}

\label{sec:results}
 
\subsection{Tidal arm properties}

\begin{figure}
\centering
\includegraphics[width=0.475\textwidth]{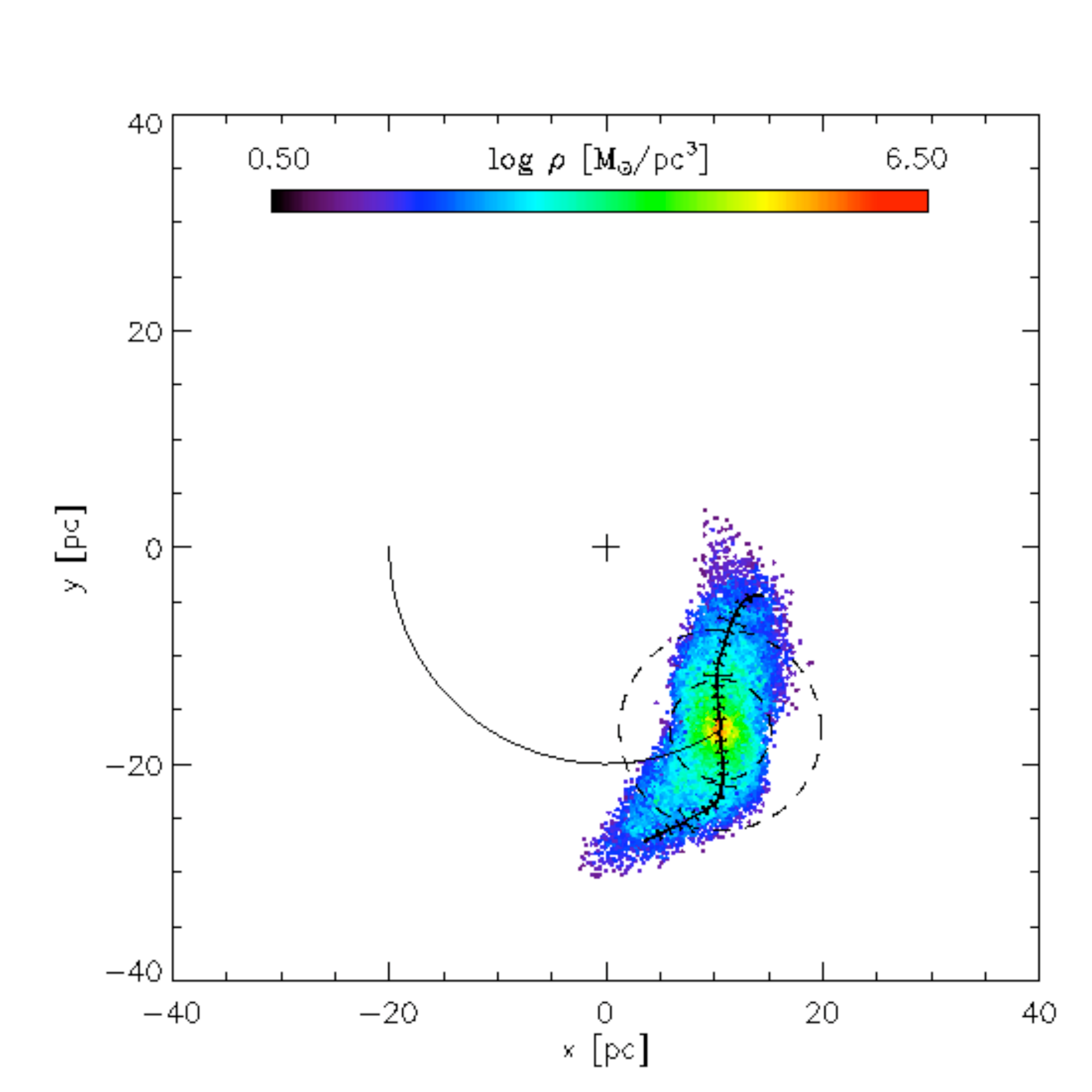} 
\caption{The model K9 at $t=0.22$ Myr. The Galactic centre is marked with a cross. The star 
cluster orbit is shown as a solid line. The dashed lines mark once and twice the membership radius.  
We look in the direction of the Galactic north pole. The short and long marks of the tidal 
arm coordinate system correspond to multiples of $1$ and $5$ pc, respectively.} 
\label{fig:armscirc100kb}
\end{figure}

\begin{figure}
\centering
\includegraphics[width=0.5\textwidth]{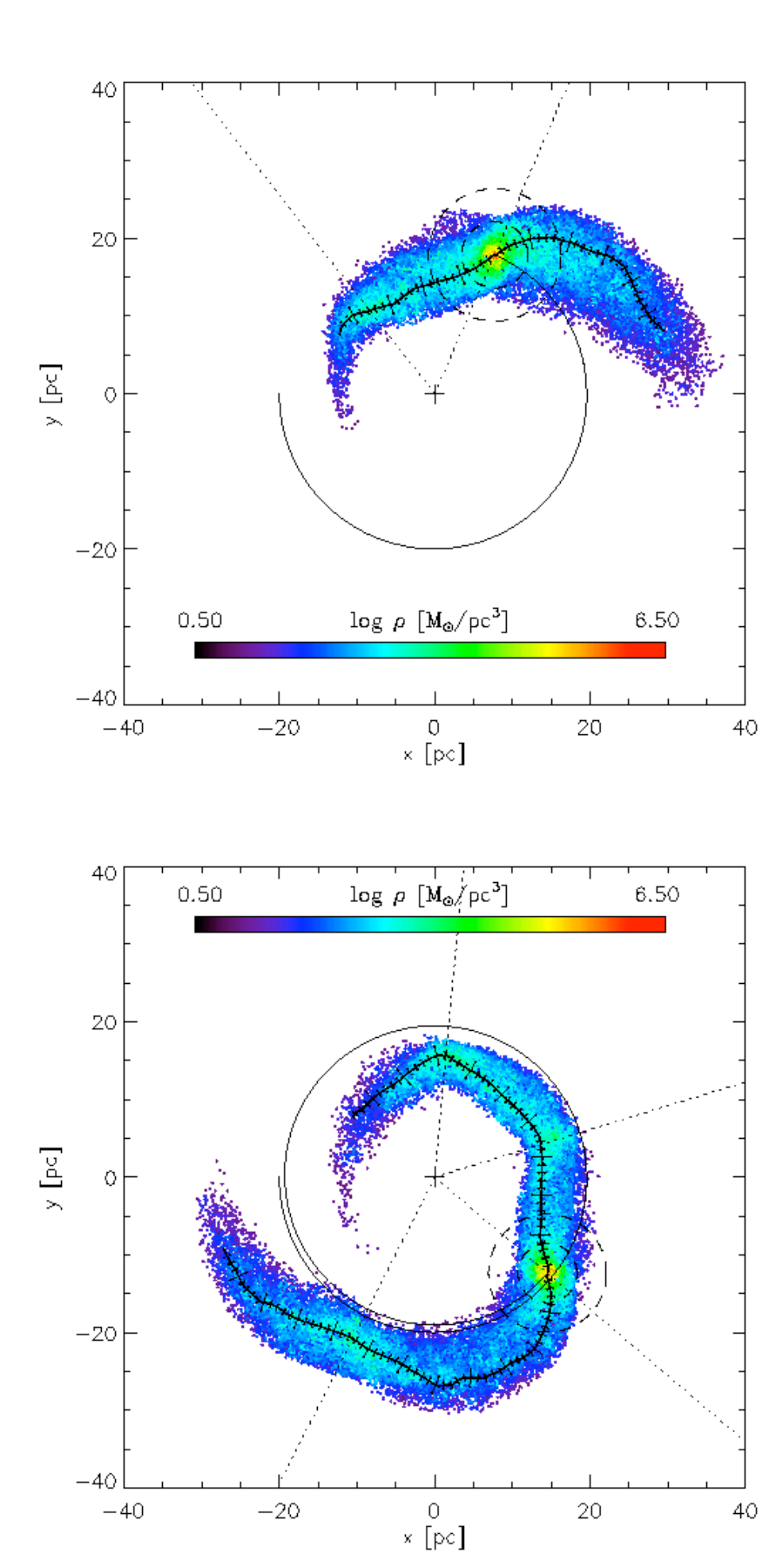} 
\caption{Further evolution of the model K9. Top panel: At $t=0.43$ Myr. Bottom panel: At $t=0.87$ Myr.
The dotted lines from the Galactic centre show the angles between density maxima
in the leading and trailing arm, respectively, with respect to the star cluster centre.} 
\label{fig:armscirc100k}
\end{figure}


\begin{figure}
\centering
\includegraphics[width=0.5\textwidth]{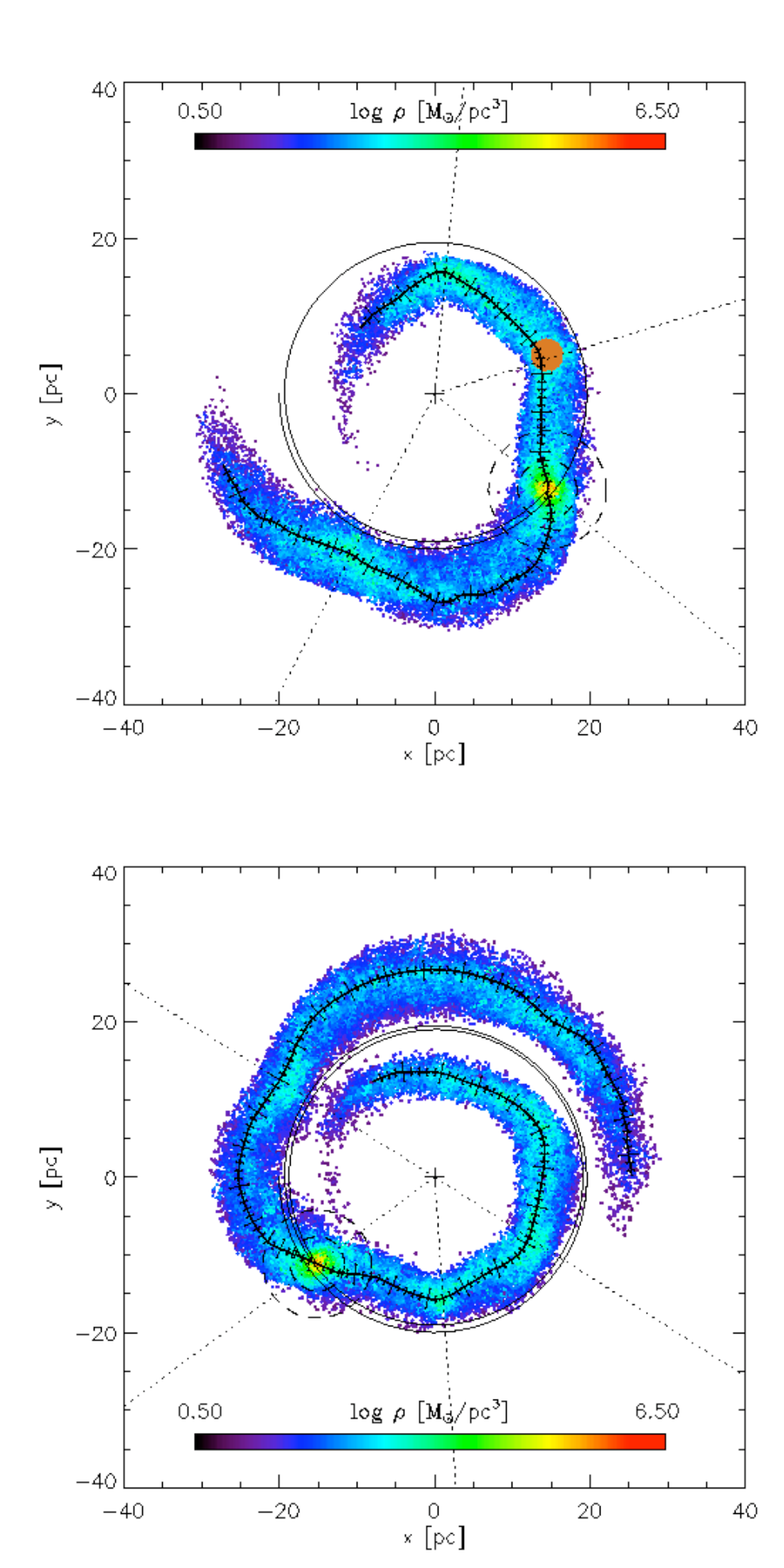} 
\caption{A standing density wave has developed in the model K9. Top panel: At $t=0.87$ Myr. A spherical cloud of tracer particles has been placed into the first clump (brown colored). 
Bottom panel: At $t=1.30$ Myr (without the tracer particles).} 
\label{fig:disperse100k}
\end{figure}

\begin{figure}
\centering
\includegraphics[width=0.5\textwidth]{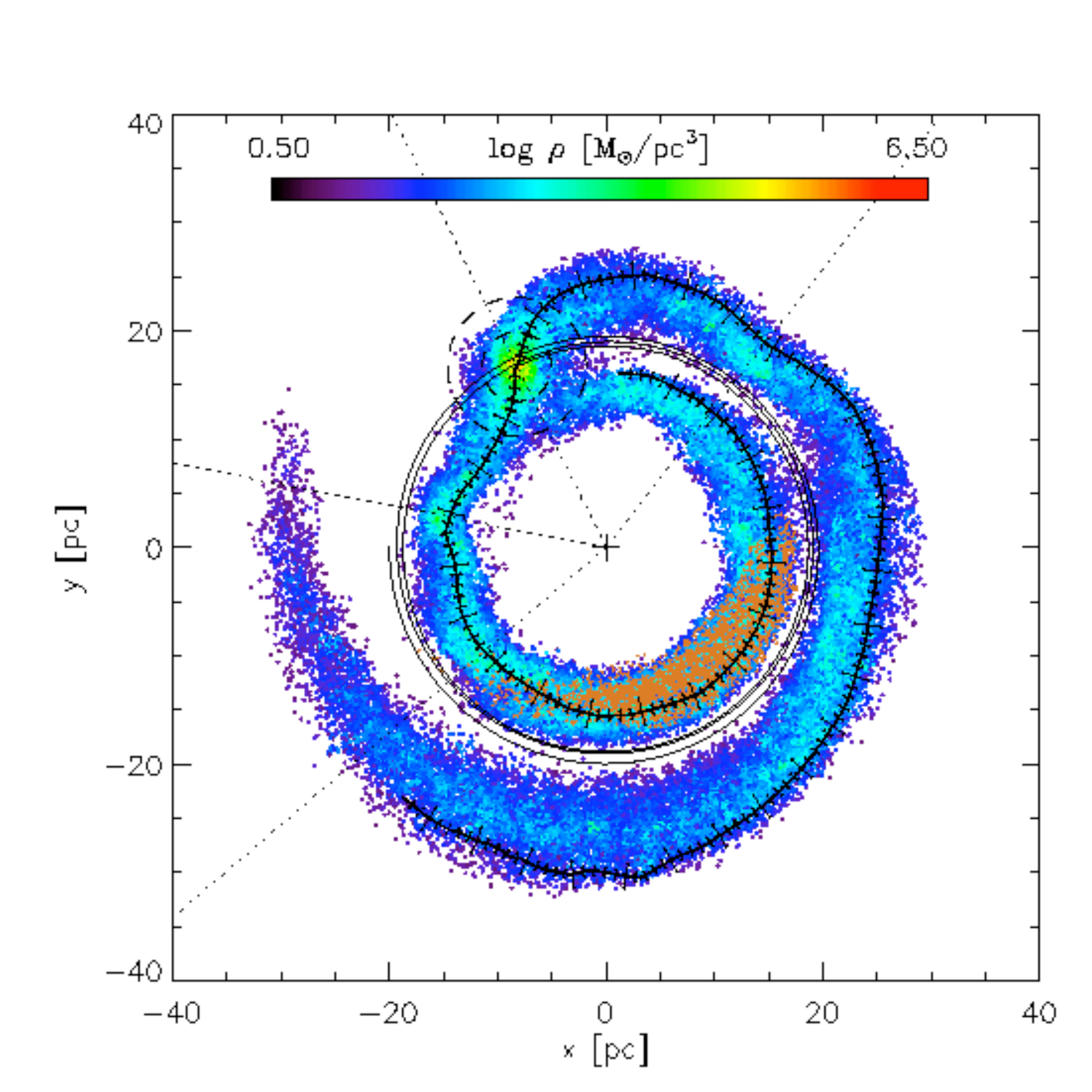} 
\caption{The model K9 at  $t=1.74$ Myr. The tracer particles have travelled further 
while the wave maximum
still persists. The leading and trailing arms have wound up. They are separated by the
potential wall of the effective potential. The tip of the leading arm has hit the remnant
of the star cluster again. We applied a weighting exponent to the particle mass in the 
expression (\ref{eq:inertiatensor}).} 
\label{fig:disperse100kb}
\end{figure}

\begin{figure}
\centering
\includegraphics[width=0.5\textwidth]{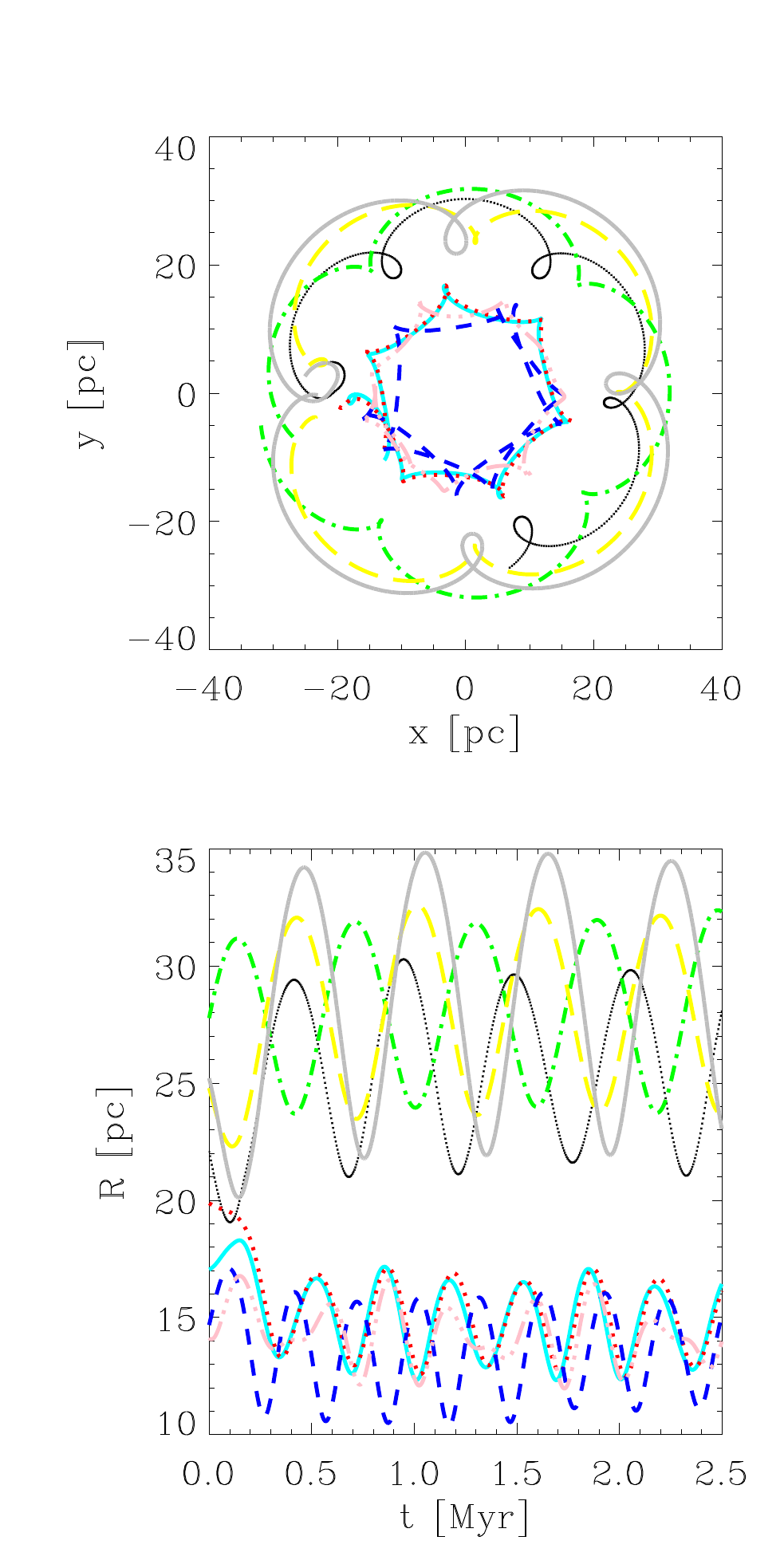} 
\caption{Top panel: Motion on cycloids within the tidal arms for a few particles in the model K9. 
The particles escape from locations near the cluster centre at (-20,0) either into the leading or the 
trailing arm. Bottom panel: Amplitude as a function of time for the same orbits. Small deviations 
from the harmonic motion can be seen which may be apparent deviations due to a slight 
change of the orbital frequency by dynamical friction or a deflection by 2-body encounters.} 
\label{fig:epi}
\end{figure}

\begin{figure}
\centering
\includegraphics[width=0.5\textwidth]{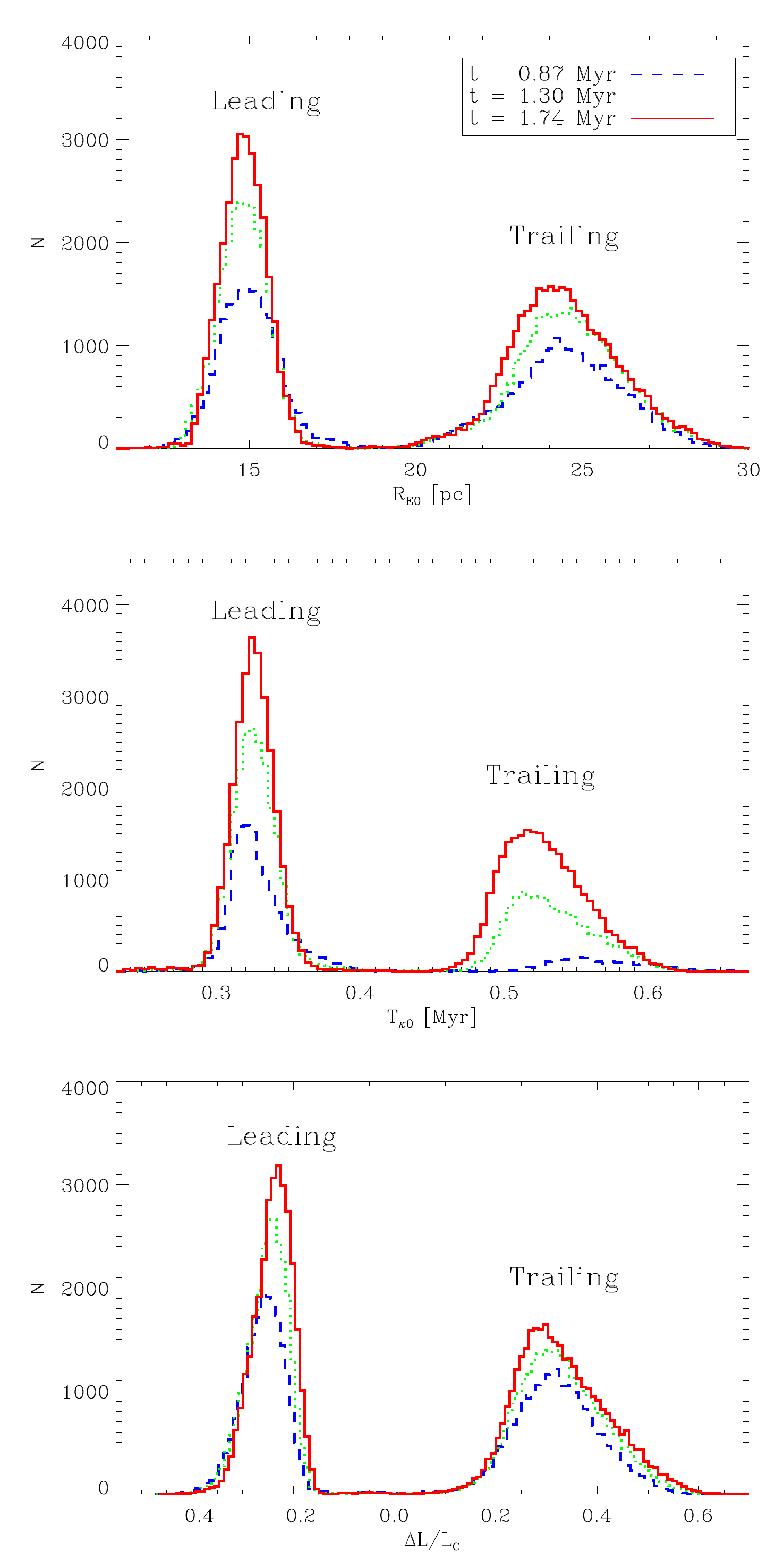} 
\caption{Histograms for the leading and trailing arms. All particles within twice (top panel)
or once (middle panel) the initial tidal radius or outside of $25$ degrees around the
density maximum in the first clump have been excluded from the statistics. 
We neglected deviations from the harmonic motion for the top and middle panel.
Top panel: Epicentre radius distribution. Middle panel: Epicyclic period distribution.
We set the bin frequency equal to the data output frequency of our code in order to avoid
an unphysical higher harmonic in the histogram.
Bottom panel: Distribution of dimensionless angular momentum differences.} 
\label{fig:histgc}
\end{figure}


\begin{figure}
\centering
\includegraphics[width=0.5\textwidth]{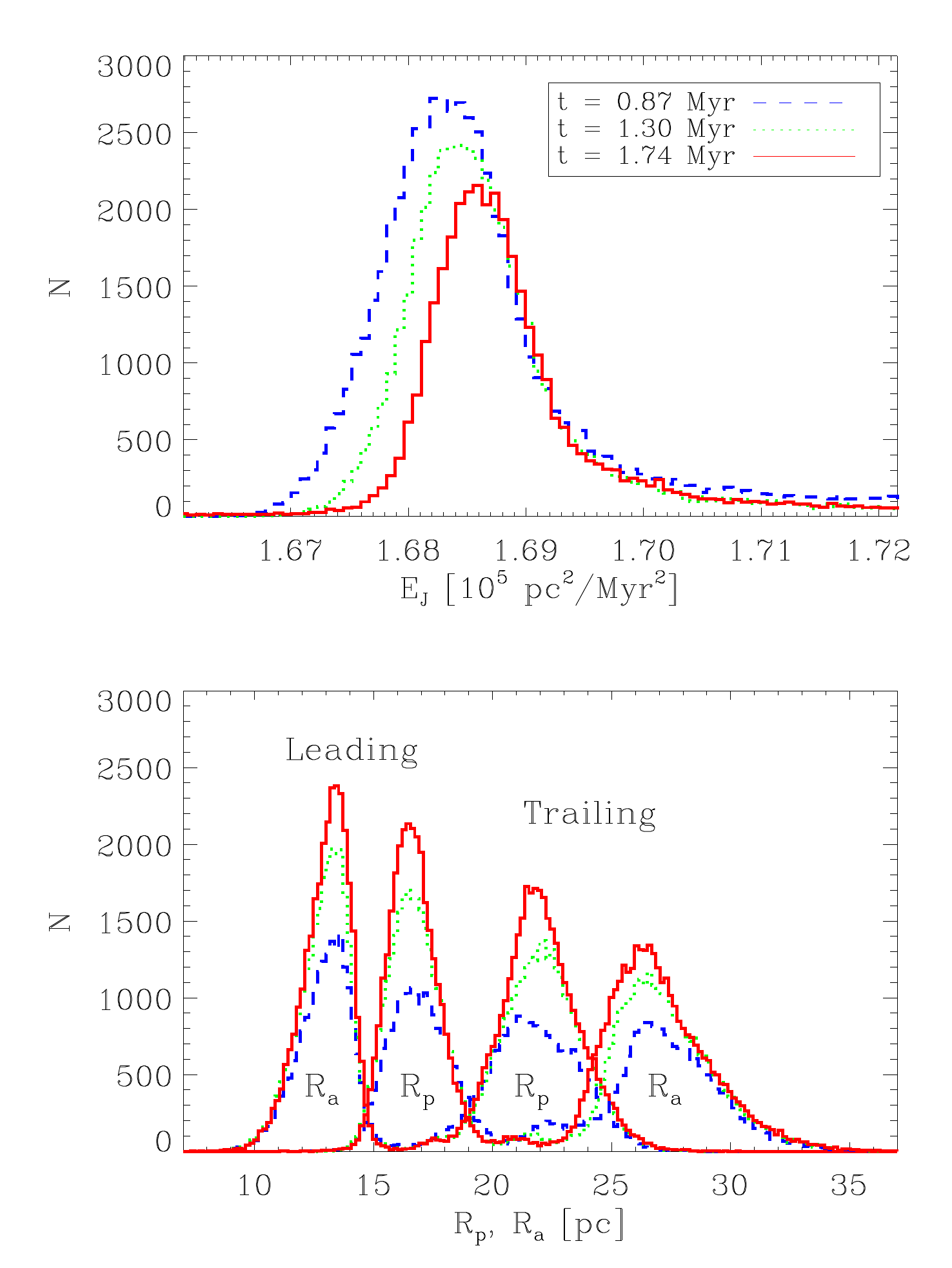} 
\caption{Top panel: Distribution of Jacobi energies. Bottom panel: Histogram
of peri- and apocentre radii of the cycloid orbits in the tidal arms.} 
\label{fig:histgc3}
\end{figure}

\begin{figure}
\centering
\includegraphics[width=0.475\textwidth]{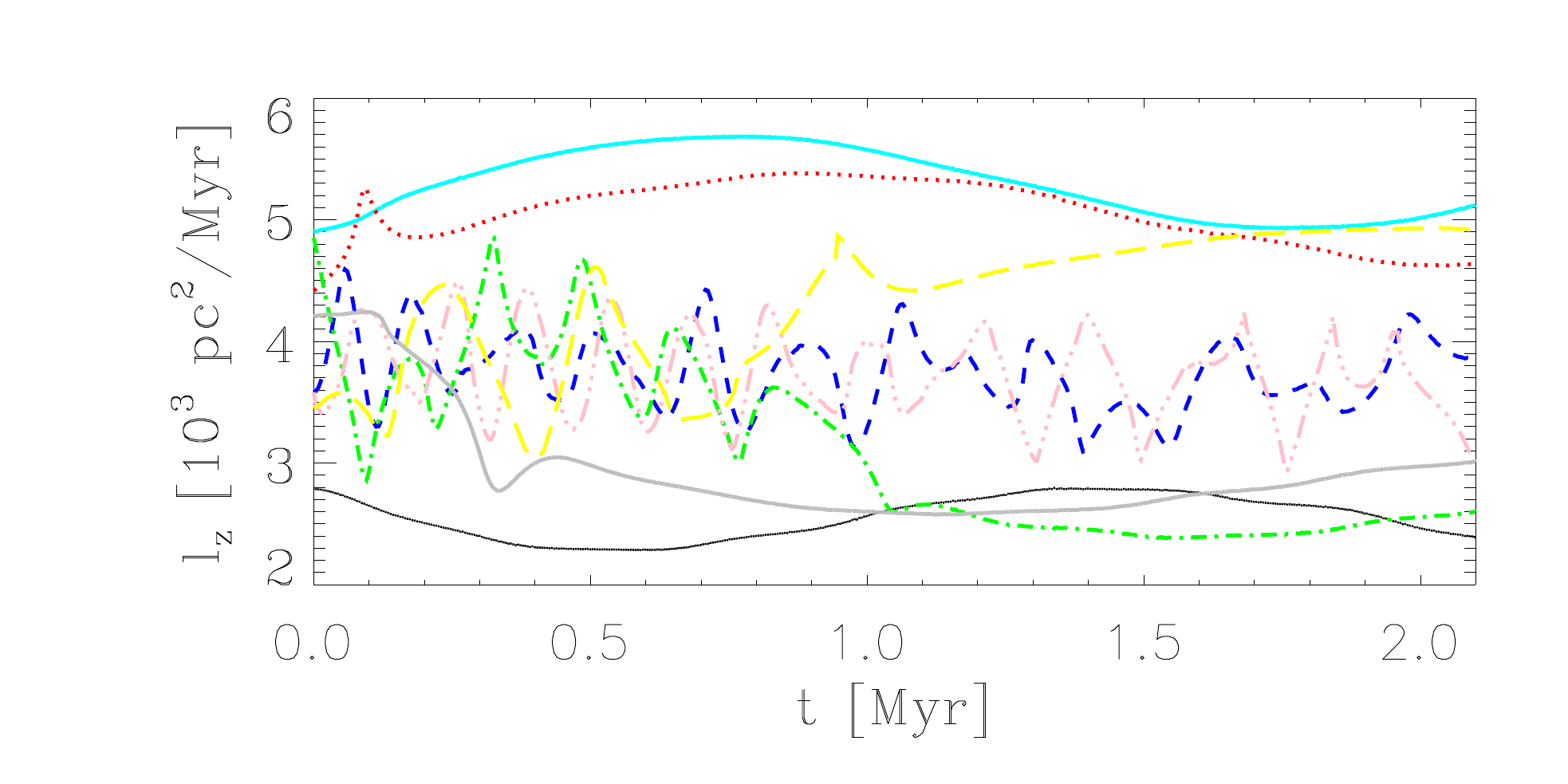} 
\caption{The $z$-component of angular momentum with respect to the Galactic centre
for a few orbits. In the cluster the angular momentum is not conserved since the
cluster potential breaks the axisymmetry of the effective potential. 
In the tidal arms the angular momentum is only approximately conserved.} 
\label{fig:amogc}
\end{figure}

Figures \ref{fig:armscirc100kb} - \ref{fig:disperse100kb} show the
formation of the tidal arms for model K9. The initial $90$\% Lagrangian radius has
been taken to be equal to the membership radius $r_m$ in Equation (\ref{eq:rtdens}). 
The star cluster dissolves in a spiral-like structure. The leading tidal arm consists of 
particles which pass the inner Lagrange point $L_1$, while the trailing arm is formed by 
particles which pass the outer Lagrange point $L_2$. The galactocentric radius $R_C$
 of the circular orbit is shown as a solid line. It decays very slowly due to dynamical friction
with the Coulomb logarithm which was modified according to Equations (\ref{eq:lnl}) 
and (\ref{eq:lnl2}). The initial value for the model K9 is $\ln\Lambda\approx 1.7$.
Most particles of the leading arm have galactocentric radii less
than $R_C$ while most particles of the trailing arm have radii larger than $R_C$.
The dashed lines mark once and twice the membership
radius $r_m$.

We have introduced a local coordinate system according to the description
in Appendix \ref{sec:tidalarmcoordinatesystem}. 
We denote the coordinate along the tidal arms as $w$, where negative
values refer to the leading arm and positive values to the trailing arm.
The short and long marks correspond to multiples of $1$ and $5$ pc, respectively.

The color coding is according to the logarithm of the stellar density.
The density clearly peaks in the cluster centre. However, one can observe clumps in 
the tidal arms where the density has local maxima. An indication for the presence
of such clumps in tidal arms  can already be found in the observations of Palomar 5 
(Odenkirchen et al. 2001, 2003). The clumps have been first noticed in simulations by 
Capuzzo Dolcetta, di Matteo and Miocchi (2005) and were investigated further by di Matteo, 
Capuzzo Dolcetta \& Miocchi (2005). They noted already the wave-like nature of this 
phenomenon.

The top panel of Figure \ref{fig:disperse100k} shows a spherical cloud of tracer particles
(coloured brown).  Figure \ref{fig:disperse100kb} shows how the tracer particles have 
travelled further into the tidal arm while the position of the density maximum in the clump 
stayed (approximately) constant. Thus the clumps can be interpreted as wave knots 
of a density wave. 
 
A theoretical explanation for such clumps was published in K\"upper, Macleod \& Heggie (2008). 
The top panel of Figure \ref{fig:epi} shows for a few particles that they move on cycloids.  The clumps appear at the position where many of the loops or turning points of the cycloids overlap. For a more
detailed theory, see Just et al. (2009). The bottom panel of Figure \ref{fig:epi} shows
the radius as a function of time. We find approximate harmonic motion in the tidal arms. 

The dotted lines from the Galactic centre in Figures \ref{fig:armscirc100k}, \ref{fig:disperse100k} and 
\ref{fig:disperse100kb} show the angles $\varphi_0$ between density maxima in the leading
and trailing arm. In order to plot these angles we determined the $w$ coordinate of the maxima 
in the mean density (cf. the top panel of Figure \ref{fig:rhosigtid} below) and obtained the corresponding Cartesian coordinates from our data files. 

Figure \ref{fig:histgc} shows the histogram of the epicentre radii $R_{E0}$ of the cycloid orbits, 
the epicyclic periods $T_{\kappa0}$, and the dimensionless angular momentum differences 
$\Delta L/L_C = (L-L_C)/L_C$ for 
different times, where $L_C$ is the angular momentum of the circular orbit. 
For the epicentre radii (and the epicyclic periods), 
stars within twice (and once) the membership radius were not included in the statistics.
The epicentre radii are given by the arithmetic mean of the last maximum and minimum
in the epicyclic amplitude. The epicyclic period is given by the time between
the last two minima in the epicyclic amplitude.
Note that at $t=0.87$ Myr not all particles have completed one epicyclic period.
For the dimensionless angular momentum differences we included only stars within 25 degrees
around the density maximum in the clump.
Thus one can see two side lobes in all panels corresponding to the leading and trailing arms. 
Figure \ref{fig:histgc3} shows the distribution of Jacobi energies $E_{J,i}$ and
the histogram of peri- and apocentre radii of the cycloid orbits in the tidal arms.

The angle $\varphi_0$  can be expressed as

\bea
\varphi_{L}  &=& \frac{2\pi}{\beta} \left[ 1-\frac{\omega_C}{\omega}\right] \\
 &=& \frac{2\pi}{\sqrt{\alpha+1}}\left[ 1 - \left( 1 + \frac{\Delta L}{L_C}\right)^{\frac{3-\alpha}{\alpha+1}}\right]  \label{eq:dllformula}
\eea

\noindent
where $\omega_C$ and $\omega$ are the circular frequencies at the radius $R_C$ 
of the circular orbit and in the vicinity of $R_C$,
$\beta$ is given by Equation (\ref{eq:beta}) and $\Delta L/L_C$ is the most frequent
dimensionless angular momentum difference. 

The subscript ``L'' refers in the
following discussion to quantities which are expressed as a function of 
$\Delta L/L_C$,\footnote{except for the case of the tidal radius $r_L$, where the subscript ``L'' refers
to the Lagrangian points $L_1$ and $L_2$}
while the subscript ``0'' refers to quantities which are directly measured
from the simulation.

The epicentre radius of the cycloids is given by

\be
R_{EL} = R_C \left( 1 + \frac{\Delta L}{L_C}\right)^\frac{2}{\alpha+1}. \label{eq:radius}
\ee

According to Just et al. (2009), the epicyclic amplitude can be 
expressed as

\be
\Delta r_L =\sqrt{ \frac{2}{\alpha + 1} \left[ \frac{3-\alpha}{2\alpha + 2} R_C^2\frac{\Delta L^2}{L_C^2}
 + \frac{\Delta E_J}{\omega_C^2}\right]} \label{eq:rm}
\ee

\noindent
to second order in the dimensionless angular momentum difference, where 
$\Delta E_J=E_J-\Phi_{\rm eff,tid}(R_C)$ is the Jacobi energy difference with respect
to the effective tidal potential at $R_C$.

Table \ref{tab:clumpangles} shows a comparison of the measured angles, epicentre radii, 
epicyclic amplitudes, peri- and apocentres and the theoretical estimates from the 
dimensionless angular momentum difference. 
We give the measured angle $\varphi_0$, the estimate
$\varphi_{L}$ according to Equation (\ref{eq:dllformula}), the error 
$\Delta\varphi/\varphi_0 = (\varphi_0 - \varphi_{L})/\varphi_0$ in percent, the
most frequent epicentre radius $R_{E0}$ in Figure \ref{fig:histgc}, the epicentre radius 
$R_{EL}$ according
to Equation (\ref{eq:radius}), the most frequent epicyclic period $T_{\kappa0}$ in 
Figure \ref{fig:histgc} and the theoretical epicyclic period $T_\kappa(R_{EL})$ using 
Equation (\ref{eq:omegau}) with the epicentre radius $R_{EL}$.
Furthermore, we give the tidal radius $r_t = \left\{ G M_{cl}/\left[(4-\beta^2)\omega_C^2\right] \right\}^{1/3}$ (King 1962), the arc length $y_0=R_C \varphi_0$, the A factors 
(Just et al. 2009),

\be
A_{y0}  = \frac{1}{\pi} \frac{\sqrt{\alpha+1}}{3-\alpha} \frac{y_0}{r_t}, \ \ \ \ \ 
A_{L} = \frac{\vert R_{E,L} - R_C\vert}{r_t}
\ee

\noindent
where $A_{y0}$ is a first-order approximation and the 
error $\Delta A/A_{y0} = (A_{y0} - A_{L})/A_{y0}$ is in percent. 
We also give the most frequent peri- and apocentre radii $R_{P0}$
and $R_{A0}$ from Figure \ref{fig:histgc3}, the most frequent scaled Jacobi energy
difference $\Delta E_J/\omega_E^2$ from Figure \ref{fig:histgc3}, the epicyclic amplitude $\Delta r_L$
from Equation (\ref{eq:rm}) and the obtained peri- and apocentre radii
$R_{PL}$ and $R_{AL}$, where  $\vert R_{PL}-R_{EL}\vert = \vert R_{AL} - R_{EL}\vert =\Delta r_L$.

There are systematic errors in both $\Phi_L$ and $A_L$ and
also in $R_{PL}$ and $R_{AL}$. The reason is shown in Figure \ref{fig:amogc}. 
In the tidal arms the angular momentum is only approximately conserved. 
The reason is the influence of the cluster potential which breaks the axisymmetry 
of the effective potential. However, in the cluster the angular momentum changes 
in a much shorter time scale. An estimate for the cumulative perturbation 
$\Delta L$ of $L$ is given by

\be
\Delta L = \Big\vert \int (\mathbf{R} \times \mathbf{a}) \, dt \Big\vert \approx \frac{R \, \Phi_{\rm cl}}{V_{\rm drift}}.
\ee

\noindent
where $\mathbf{a}$, $R=\vert\mathbf{R}\vert$, $\Phi_{\rm cl}$ and
$V_{\rm drift}$ are the acceleration, galactocentric radius, potential energy of the cluster and 
the drift velocity, respectively. Thus a slow drift velocity increases the change in $L$. 
Here a more detailed theory is desirable.
 
 \begin{table*}
\begin{center}
\begin{tabular}{ccccccccccccccc}
\hline
\hline
\# & $t$ [Myr] & Arm & Clump &  $\Delta L/L_C$ & $\vert\varphi_0\vert$ [deg.] & $\vert\varphi_{L}\vert$ [deg.] & $\Delta\varphi/\varphi_0$ [\%] & $R_C(t)$ [pc]  \\
\hline
1 & 0.87 & lead. & 1 & -0.232 & 57.5 & 47.1 & 18.1 & 19.0   \\
2 &    "     & lead. & 2 & -0.273 & 67.6 & 55.7  & 17.6 & "  \\
3 &   "      & trail. & 1 & 0.320 & 68.6  & 61.9 & 9.8 & "   \\
4 & 1.30 & lead. & 1& -0.215 & 57.5 & 43.6 & 24.2  & 18.8  \\ 
5 &    "     & lead. & 2 & -0.227 & 53.3 & 46.1 & 13.5 &    "    \\
6 &    "     & trail. & 1 & 0.276 & 68.4 & 53.6  & 21.6  &      "  \\
7 & 1.74 & lead. & 1 & -0.209 & 52.6 & 42.4 & 19.4  & 18.6 \\ 
8 &     "     & lead. & 2 & -0.205 & 51.5 & 41.5 & 19.4  &    "      \\
9 &     "    & trail. & 1 & 0.273 & 64.2 &  53.0 & 17.4  &      "  \\
\hline
\hline
\# & $R_{E0}$ [pc]  & $R_{EL}$ [pc]  & $T_{\kappa0}$ [Myr] & $T_\kappa(R_{EL})$ [Myr] & $r_t (t-T_{\kappa0})$ [pc]  & $y_0$ [pc] & $A_{y_0}$ & $A_{L}$\\
\hline
1 & 14.9 & 14.9 & 0.32 & 0.35    & 2.67 & 19.1 & 1.88 & 1.54   \\
2 & " & 14.2 &     "    &  0.34  &     "    &  22.4 & 2.20 & 1.80   \\
3 & 24.2 &  24.5 & 0.55 & 0.55 & 2.87 & 22.7 & 2.07 & 1.92   \\
4 & 14.7 & 15.1 & 0.33 & 0.36  & 2.40 & 18.9 & 2.07 & 1.54  \\
5 &    "  & 14.9 &     "    &  0.35  &     "    & 17.5 & 1.91& 1.63  \\
6 & 24.7 & 23.5 & 0.51 & 0.53 & 2.63 & 22.4 & 2.23 & 1.79  \\
7 & 14.7 & 15.0 & 0.32 & 0.36  & 2.21 & 17.1 & 2.03 & 1.63 \\
8 &    "  & 15.1 &      "    &  0.36  &     "    & 16.7 & 1.98 & 1.58  \\
9 & 24.0 & 23.2 &  0.51 & 0.53 & 2.29 & 20.8 & 2.38 & 2.01   \\
\hline
\hline
\hline
\# & $A_{R_{E0}}$ & $\Delta A/A_{y0}$ [\%] & $R_{P0}$ [pc] & $R_{A0}$ [pc] & $\Delta E_J/\omega_E^2$ [pc$^2$] & $\Delta r_L$ [pc] & $R_{PL}$ [pc] & $R_{AL}$ [pc] \\
\hline
1 & 1.54 & 18.1 & 16.6 & 13.6 & 3.80 & 2.8 & 17.7 & 12.1 \\
2 &    "    &  18.1 &  "  &   "          & 3.46 & 2.9 & 17.1 & 11.3 \\
3 & 1.81 & 7.2 & 21.3 & 26.2   & 9.25 & 5.3 & 19.2 & 29.8 \\
4 & 1.71 &  25.6 & 16.5 & 13.3 & 6.50 & 3.1 & 18.2 & 12.0 \\
5 &    "    &  14.7 &  "   &    "        &  6.39 & 3.1 & 18.0 & 11.8 \\
6 & 2.24 & 19.7 & 22.1 & 26.2 &  14.4 & 5.1 & 18.4 & 28.6 \\
7 & 1.76 & 19.7 & 16.5 & 13.3 & 11.5  & 3.7 & 18.7 & 11.3 \\
8 &    "    & 20.2 &  " &    "          &  11.6 & 3.7  & 18.8 & 11.4 \\
9 & 2.36 & 15.5 & 21.4 & 26.0 &  25.2 & 6.0 & 17.2 & 29.2 \\
\hline
\end{tabular} 
\end{center}
\caption{Comparison of measurement and theory for the angles of the density maxima and 
the A factors. For explanations see the text.} 
\label{tab:clumpangles}
\end{table*}

\begin{figure}
\centering
\includegraphics[width=0.5\textwidth]{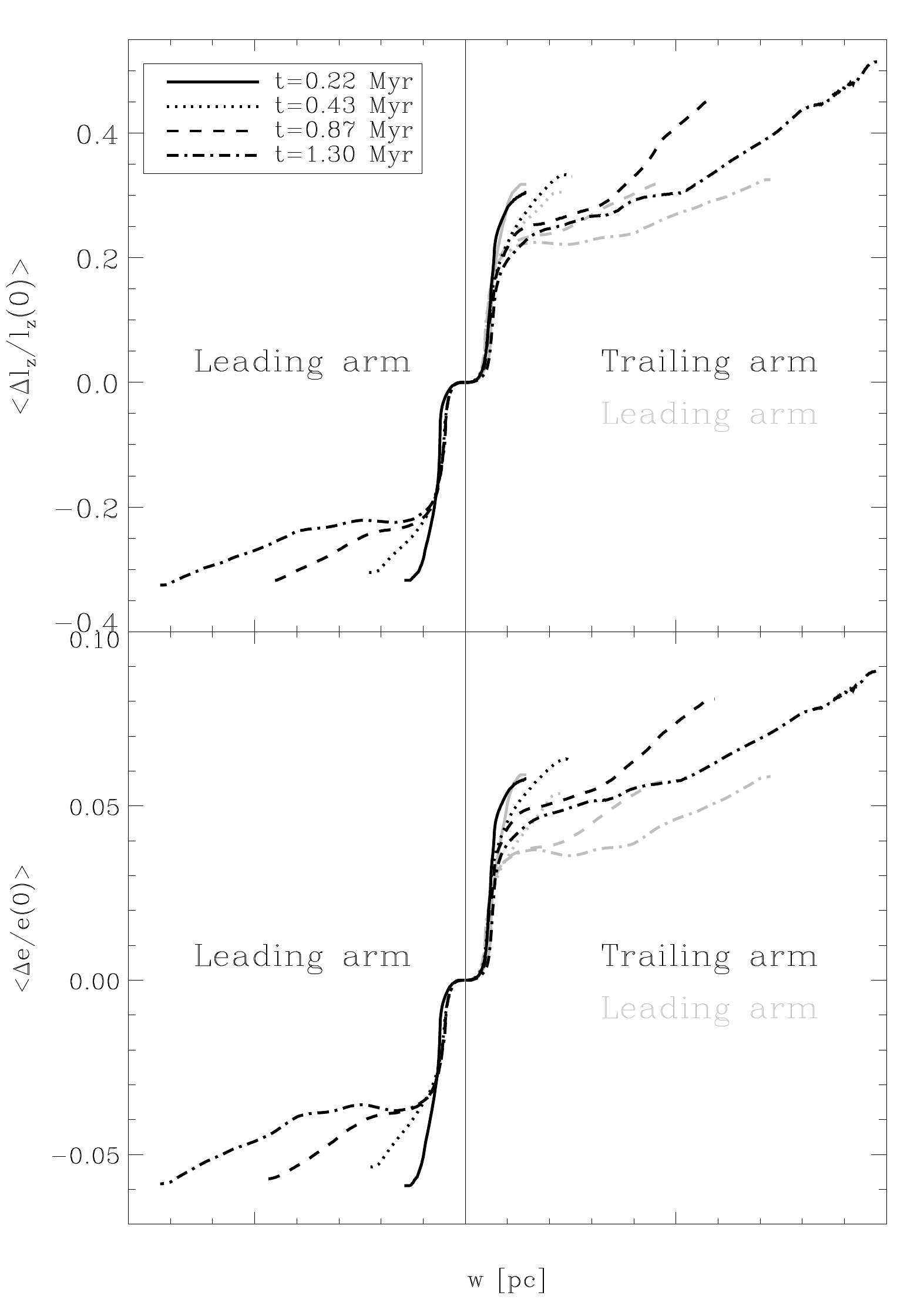} 
\caption{Top panel: Time evolution of the angular momentum difference between the cluster center and
the position in the tidal arm for the model K9. The angular momentum difference is normalized by the
angular momentum of the cluster center.
 In order to show the asymmetry between the leading and trailing arms,  
the lines for the leading arm have been rotated by 180 degrees about the origin and replotted 
in grey.
Bottom panel: As in the top panel, but for the energy (internal and external) difference between the 
cluster center and the position in the tidal arm for the model K9. The energy difference is normalized 
by the energy of the cluster center.} 
\label{fig:amoentid}
\end{figure}

\begin{figure}
\centering
\includegraphics[width=0.5\textwidth]{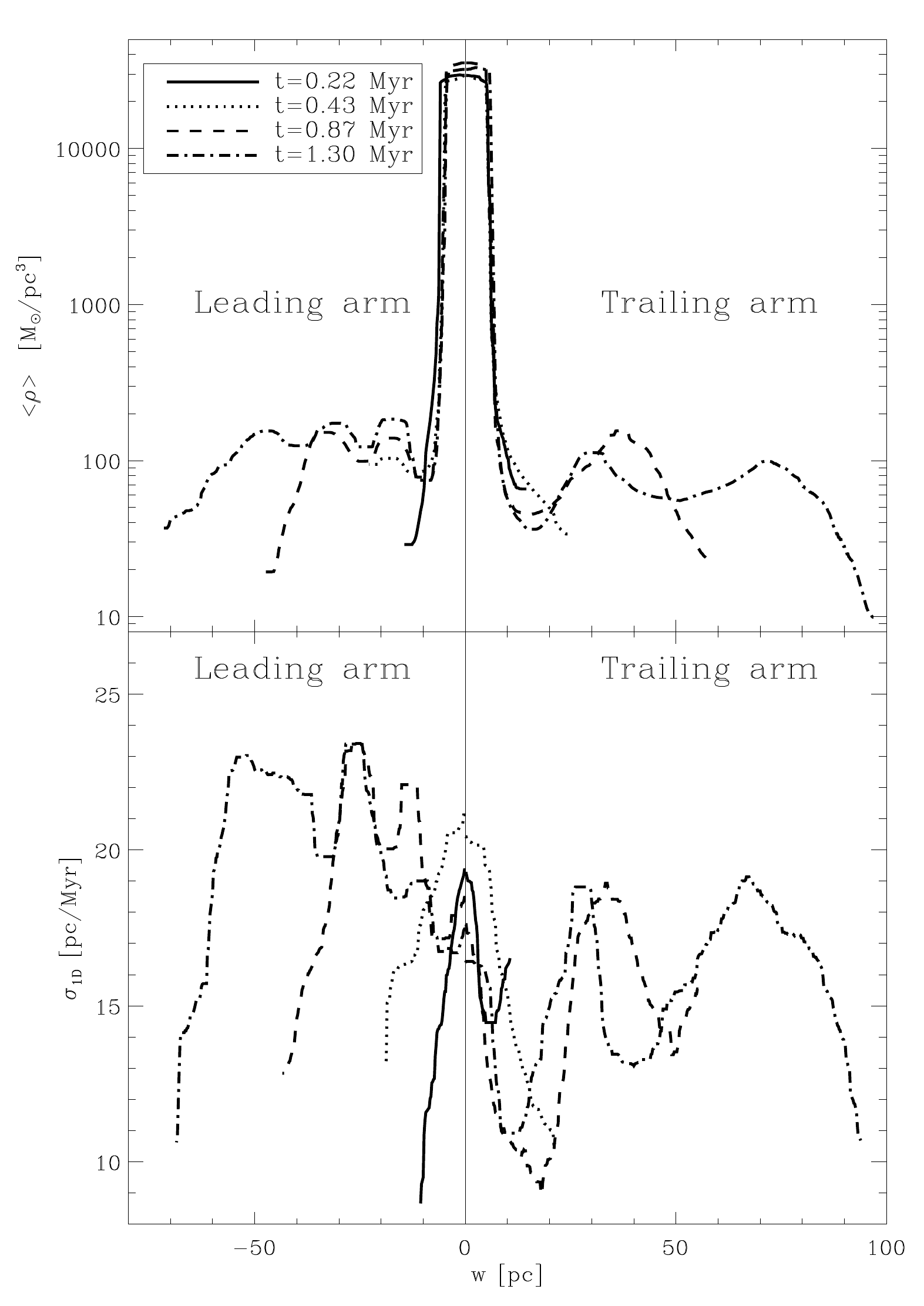} 
\caption{Top panel: Time evolution of the mean density of stars along the tidal arms 
for the model K9. One can see that several density wave maxima develop with time.
Bottom panel: Time evolution of the 1D velocity dispersion along the tidal arms for the model K9.
 The characteristic features in the mean density (top panel) can also be seen in the 
 velocity dispersion.} 
\label{fig:rhosigtid}
\end{figure}

\begin{figure}
\centering
\includegraphics[width=0.5\textwidth]{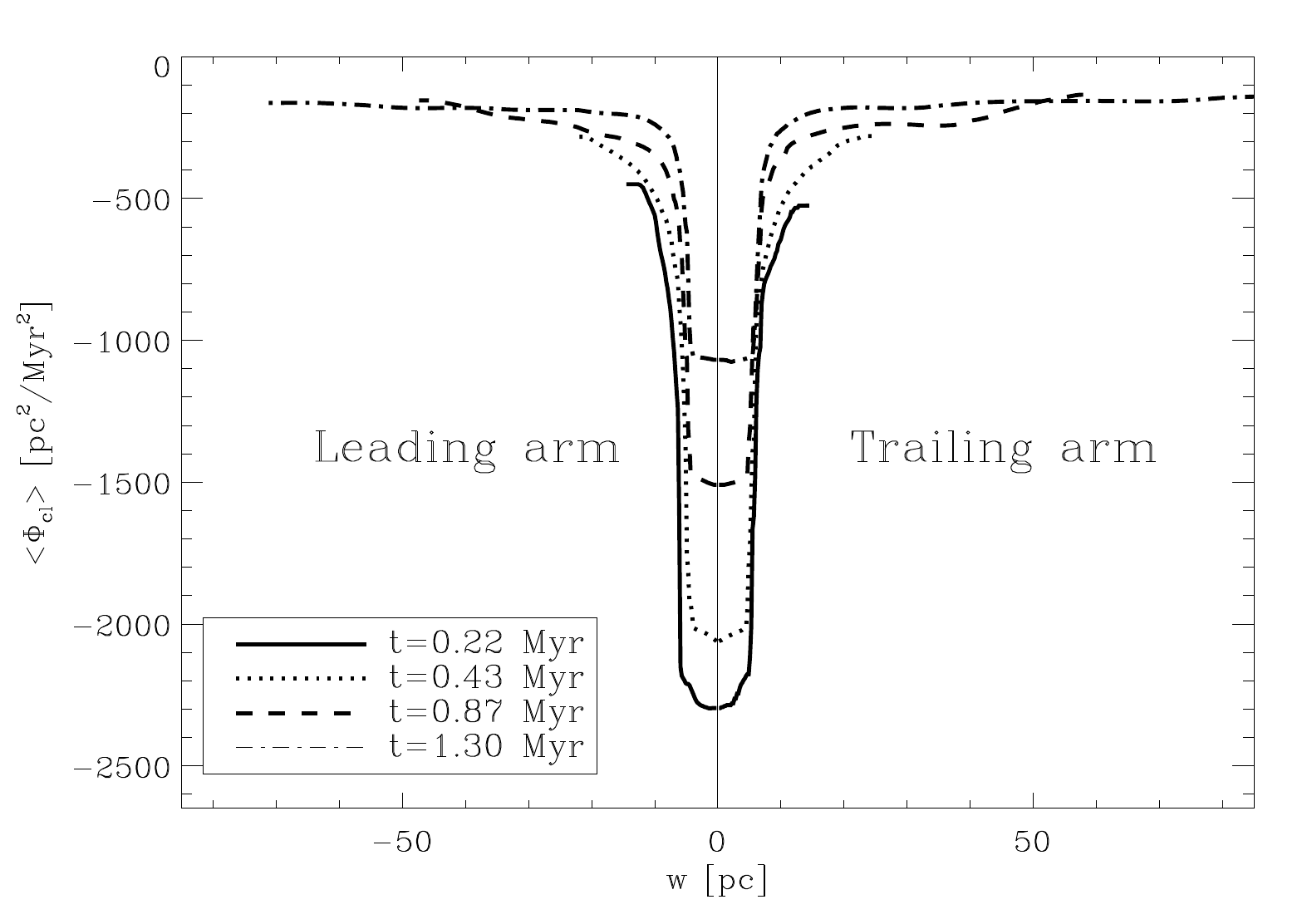} 
\caption{Time evolution of the mean star cluster potential  
along the tidal arms for the model K9. One can see that the potential well is deeper in the beginning but gets
shallower as the cluster loses mass.} 
\label{fig:phitid}
\end{figure}

For Figures \ref{fig:amoentid} - \ref{fig:phitid}, we averaged over spheres with a radius which was
approximately equal to the width of the tidal arms in the $xy$ plane (see Appendix \ref{sec:tidalarmcoordinatesystem} for the details). 
Since the Figures were still noisy, we have used a median smoothing in addition. 
The width of the smoothing kernel has been taken to be twice 
the membership radius $r_m$ defined in Equation (\ref{eq:rtdens}).

For four different times, the top panel of Figure \ref{fig:amoentid} shows the $z$-component of the 
dimensionless mean angular momentum difference $\left[l_{z}(w) - l_{z}(0)\right]/l_{z}(0)$ 
along the tidal arms. 
The specific angular momentum was calculated with respect to the Galactic center. 
In order to show the asymmetry between the leading and trailing arms, the lines for 
the leading arm have been rotated by 180 degrees about the origin and replotted in grey.
This kind of asymmetries arise from the geometry of the effective potential.

The bottom panel of Figure \ref{fig:amoentid} shows the same for the mean 
energy difference $\left[e(w)-e(0)\right]/e(0)$.
The specific energy was calculated with respect to the Galactic center. 
For the definition of the 
energy, see Section \ref{sec:energycheck}. 
A positive energy difference corresponds to the trailing arm while a negative 
energy difference corresponds to the leading arm.
This is in accordance with the positive normalization in Equation (\ref{eq:phi0scalefree}) 
for the scale free potential in Equation (\ref{eq:phiscalefree}).
This Figure also shows an asymmetry between leading and trailing
arm. What can be seen in this plot is that with the time more and more particles with a low 
energy difference with respect to the cluster center stream into the tidal arms.
Thus the modulus of the mean energy difference falls off with time. 
In this connection it is worthwhile to mention that the particles with a low Jacobi energy 
stream into the tips of the tidal arms. This can be seen in Figure \ref{fig:disperse100kb}: 
The stars in the tips of the tidal arms are far away from the
potential wall of the effective tidal potential which lies below the solid line of the orbit
of the star cluster center.

The top panel of Figure \ref{fig:rhosigtid} shows the density profile along the tidal arm coordinate $w$. 
At $t=0.43$ Myr, one clump can be seen in the leading arm. At $t=0.87$ Myr,
two clumps can be seen in the leading arm and one in the trailing arm.
At $t=1.30$ Myr, three clumps can be seen in the leading arm and two in the
trailing arm.

The bottom panel of Figure \ref{fig:rhosigtid} shows the profile of the 1D velocity dispersion 
along the tidal arm coordinate $w$. The averaging for the calculation of the velocity
dispersion has been done in small spheres around the tidal arm coordinate system
whose radius was approximately 1/5 of the width of the tidal arms in the $xy$ plane 
(see Appendix \ref{sec:tidalarmcoordinatesystem} for more details).
The velocity dispersion profile also exhibits local 
maxima at the positions of the density maxima. This is in accordance with the notion that the
clumps in the tidal arms occur at the positions where many of the loops or turning
points of the cycloid orbits overlap. At these positions, the random velocities
should exhibit maxima as well. 
Note that the first maximum in the leading arm at 
$t=0.43$ Myr cannot yet be seen clearly in the bottom panel of Figure \ref{fig:rhosigtid}. Figure 
\ref{fig:armscirc100k} shows that this density maximum is still
in the process of building up.

Figure \ref{fig:phitid} shows the profile of the cluster gravitational potential 
along the tidal arm coordinate $w$ at four different times. One can see that the potential 
well of the star cluster is deeper in the beginning but gets shallower as the cluster loses mass.

\begin{figure}
\centering
\includegraphics[width=0.5\textwidth]{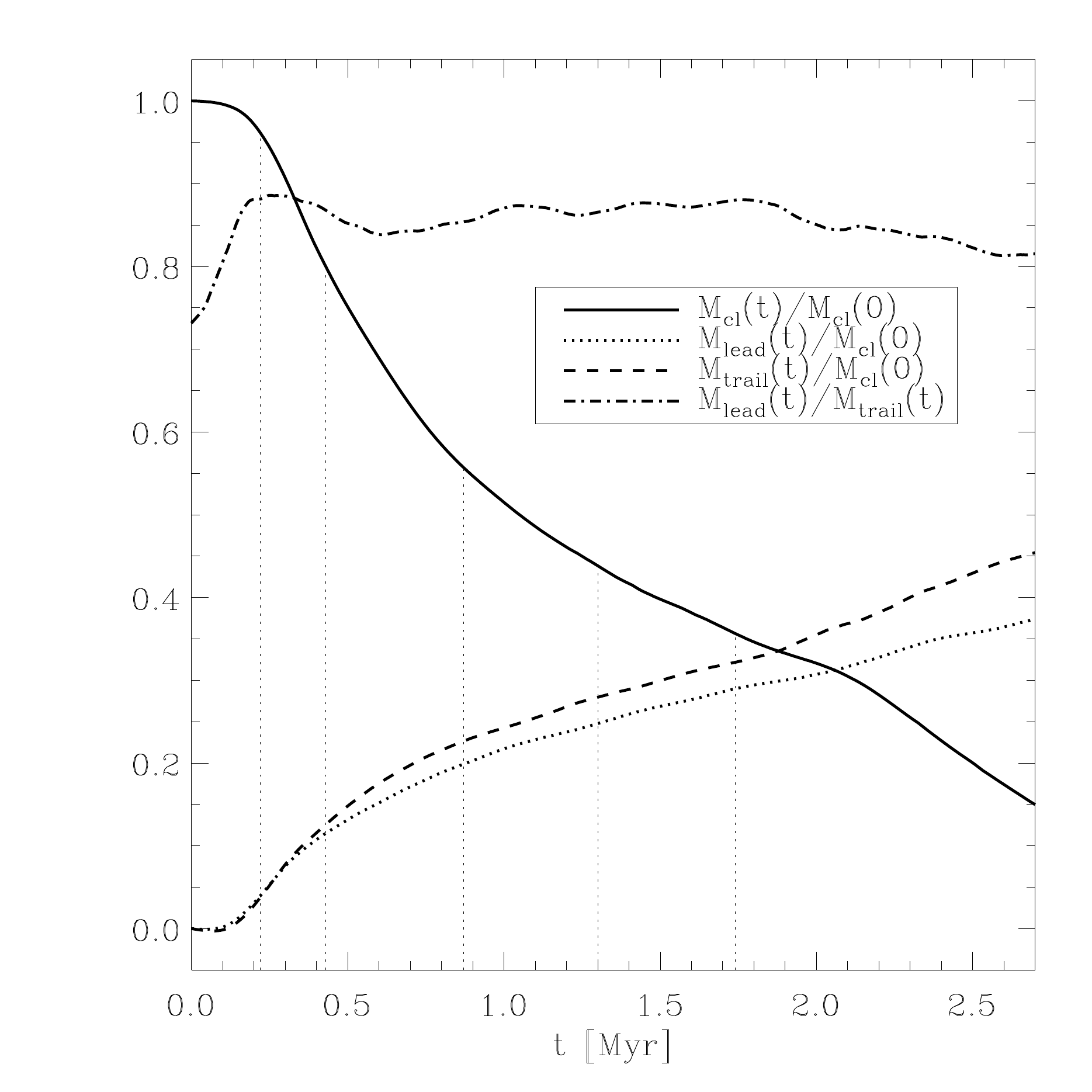} 
\caption{Evolution of the cluster mass within the tidal radius and
the mass in the tidal arms for the model K9. It can be seen that more particles escape into the
trailing arm than into the leading arm. The ratio of leading arm mass to trailing arm mass
is always roughly 85\%. The thin vertical dotted lines correspond to $t=0.22, 0.43, 0.87, 1.30$
and $1.74$ Myr.}
\label{fig:escaper3}
\end{figure}
 
Figure \ref{fig:escaper3} shows the evolution of the cluster mass contained within the
tidal radius for the model K9. In addition, the mass in the tidal arms is shown as a function of time.
It can be seen that more particles escape into the trailing arm than into the
leading arm. In the relaxation-driven dissolution scenario this would be paradoxical
since the inner Lagrange point $L_1$ is at a lower energy than the outer Lagrange 
point $L_2$ according to Figure \ref{fig:gczoom2d}. 
However, most stars are in the high-energy region of the star cluster. For these particles
the phase space for escape into the trailing arm is larger than that for
escape into the leading arm.

\subsection{Lifetime scaling and RE classification}

\label{sec:lifetimescaling}

\begin{figure}
\centering
\includegraphics[angle=90,width=0.5\textwidth]{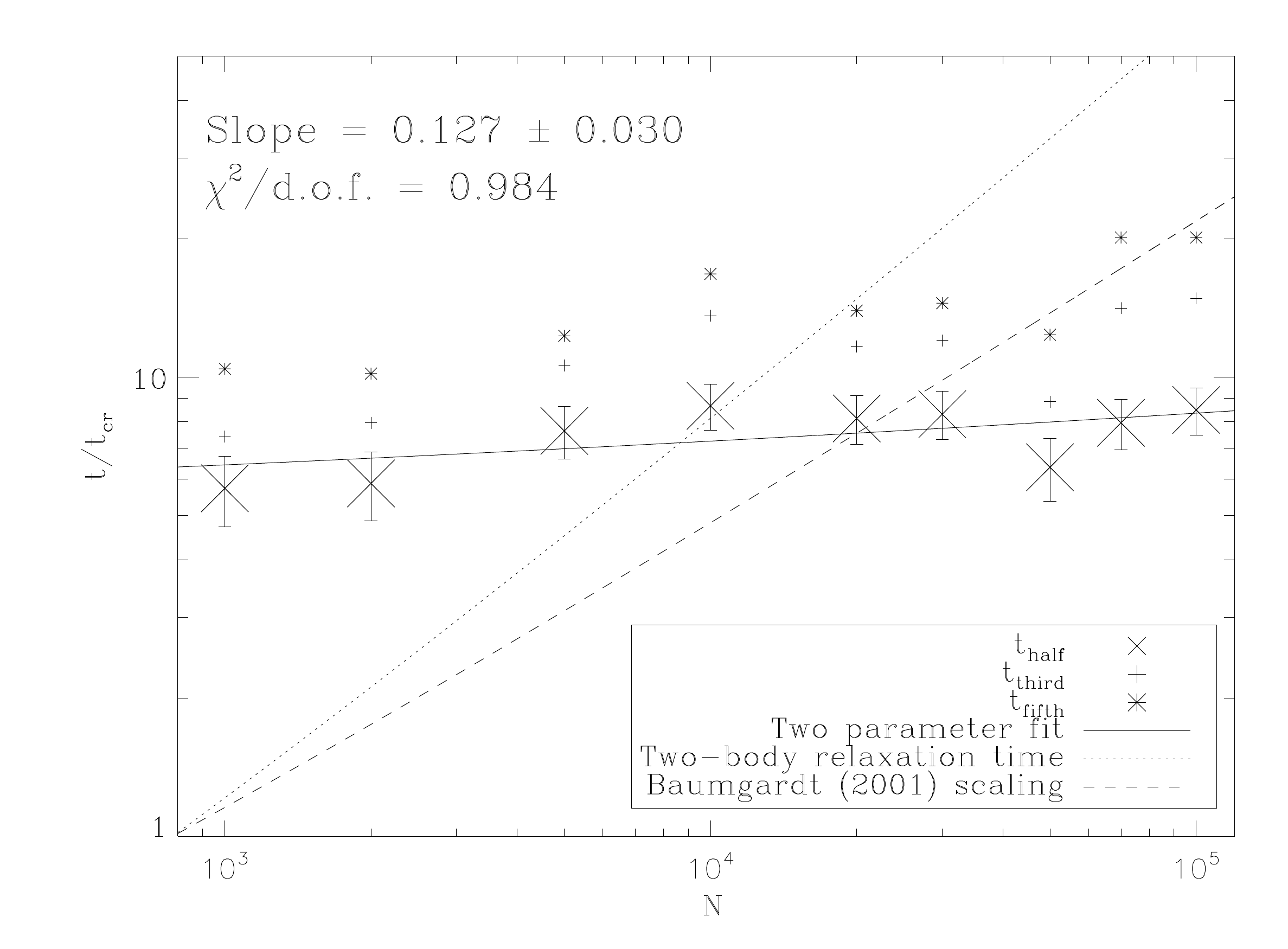} 
\caption{Scaling of the ratio of half-mass time $t_{half}$ to crossing time $t_{cr}$ 
as a function of the particle number $N$ for the models K1 - K9. The errors correspond to $1$ crossing 
time $t_{cr}$. We have $t_{cr}= 0.123$ Myr in all models. The ``third-'' and ``fifth-mass'' times
are also shown.} 
\label{fig:baumgardt}
\end{figure}

\begin{figure}
\centering
\includegraphics[width=0.5\textwidth]{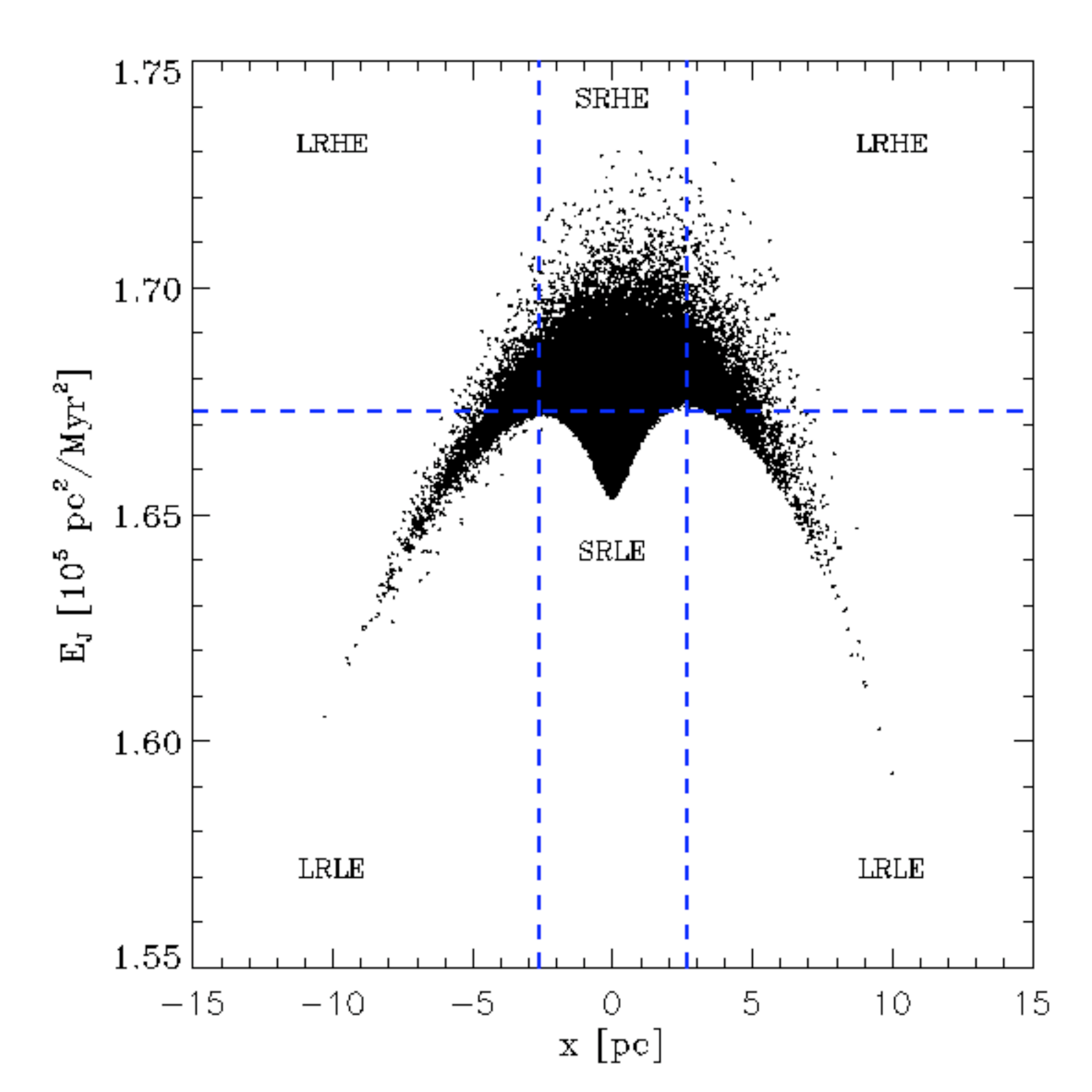} 
\caption{Projection of initial Jacobi energies of stars onto the $x-\Phi_{\rm eff}$ plane of Figure
\ref{fig:gczoom2d} for the model K9. The dashed lines mark the different regions of the
Radius-Energy (RE) classification.} 
\label{fig:jacobi2}
\end{figure}

Figure \ref{fig:baumgardt} shows the scaling of the half-mass time as a function of the 
particle number. The half-mass time is the time after which the star cluster has lost half of its 
initial mass due to escaping stars. It also shows as dots the times, when the cluster has one 
third or one fifth, respectively, of its initial mass.
The particle number in Figure \ref{fig:baumgardt} ranges from $N=10^3$ up to $N=10^5$. 
The $\chi^ 2$ fit of a power law shows that the half-mass time
depends only slightly on the particle number $N$. 

This is in contrast to the
relaxation-driven dissolution of star clusters. If the dissolution is relaxation-driven, 
the stars are scattered above the escape energy (or the critical Jacobi energy) by 
two-body relaxation before they can escape through exits in the equipotential 
surfaces around the Lagrangian points $L_1$ and $L_2$. 
In this case the half-mass or dissolution time should depend more strongly on the particle number
than in Figure \ref{fig:baumgardt}. 
Baumgardt (2001) developed a detailed theory  for the relaxation-driven
dissolution of star clusters on circular orbits in a steady tidal field with back-scattering of potential escapers in which the half-mass time scales as $t_{half} \propto t_{rh}^{3/4}$. 

Fukushige \& Heggie (2000) give a hint for the understanding of the scaling of the 
half-mass time in our models. In Figure \ref{fig:jacobi2} one can see
that the cluster fills the energetic region above the total effective potential of 
Figure \ref{fig:gczoom2d}. 
Many stars are initially outside of the tidal radius 
$\overline{r}_L = \left[x(L_1)+x(L_2)\right]/2$ and most particles have Jacobi energies $E_J$ 
per unit mass which are higher than the mean effective potential 
$\overline{E}_{J,L} = \left[E_J(L_1)+E_J(L_2)\right]/2$ of 
both Lagrange points $L_1$ and $L_2$. We initially have for the model K9 the following ratio of 
particle numbers:

\be
\frac{N_{\rm r > \overline{r}_L}}{N_{\rm r < \overline{r}_L}} \approx 0.37 \ \ \ \mathrm{and} \ \ \ 
\frac{N_{\rm E_J > \overline{E}_{J,L}}}{N_{\rm E_J < \overline{E}_{J,L}}} \approx 1.80.
\ee

\noindent
The stars which are outside of $\overline{r}_L$ and the high-energy particles with respect to the 
critical Jacobi energy $\overline{E}_{J,L}$ can leave the cluster relatively fast as compared with 
the relaxation time, provided they are not bound by a non-classical integral of motion which would 
hinder their escape. Figure \ref{fig:poingc2} shows that above a certain Jacobi 
energy threshold in the high-energy regions all orbits are chaotic and not subject to a 
non-classical integral of motion.

It is possible to classify the particles initially according to their membership to one of four regions:

\begin{enumerate}
\item Large Radius High Energy (LRHE) region
\item Small Radius High Energy (SRHE) region
\item Large Radius Low Energy (LRLE) region
\item Small Radius Low Energy (SRLE) region
\end{enumerate} 

\noindent
The distinction between these regions is shown with dashed lines in Figure \ref{fig:jacobi2}.
We call this the Radius-Energy (RE) classification. The classification arises 
due to the existence of the Lagrange points $L_1$ and $L_2$ at proximate (or equal)
Jacobi energies. For the model K9 with $N=10^5$, we initially have the following 
occupation numbers of the
four regions,

\bea
&&N_{\rm LRHE} = 24370, \ \ \ N_{\rm SRHE} = 39854, \nonumber \\ 
&&N_{\rm LRLE} = 2831, \ \ \ N_{\rm SRLE} = 32945. \label{eq:occupationnumbers}
\eea

\noindent
In an exact treatment the critical equipotential line should be taken as the dividing 
line between small and large radii in the RE classification. However, in real $N$-body simulations 
it is more convenient to adopt the tidal radius for this purpose as has been done
in the counting for Equations (\ref{eq:occupationnumbers}). 
Particles in the LRLE region
are immediately lost due to the energy barrier if they are not bound to the cluster by a
non-classical integral of motion. Particles in the LRHE and SRHE
regions can mix dependent on their individual position and velocity. Particles in the
SRLE region are bound to the cluster until they are lifted to the SRHE region by
secular evolution or 2-body relaxation.

Particularly the ratio

\be
\alpha_M = \frac{M_{\rm LRLE} + M_{\rm LRHE} + M_{\rm SRHE}}{M_{\rm SRLE}}
\ee
 
\noindent
determines the physics of the dissolution process, where $M_{\rm LRLE}$, 
$M_{\rm LRHE}$, $M_{\rm SRHE}$ and $M_{\rm SRLE}$ are the occupation masses
of the four regions.
If $\alpha_M$ is close to zero the main process which leads to the dissolution of the cluster is two-body 
relaxation, which scatters stars from the SRLE region into the two high-energy regions. The larger
the particle number $N$ is, the slower is this process. We speculate that
$\alpha_M$ was very small for the old globular clusters in the halo 
of the Milky Way and that their dissolution is relaxation-driven, but that many 
young star clusters (open clusters or associations) with larger values of $\alpha_M$ may form at 
all times in the Milky Way and dissolve fast as compared with the Hubble time. 
If $\alpha_M$ is sufficiently large, mass loss from the SRLE region seems to be 
dominated by a self-regulating process of increasing Jacobi energy
due to the weakening of the potential well of the star cluster, which is 
induced by the mass loss itself (Just et al. 2009).
A simple estimation shows that the critical Jacobi energy $\overline{E}_{J,L}$ increases more
slowly with time as compared with the Jacobi energy $E_J$ of a star in the
non-stationary gravitational potential of the star cluster.
While the particles in the LRLE, LRHE and SRHE regions of the  star cluster move away from
the cluster along the tidal arms, particles are continually shifted from the SRLE region into 
the two high-energy regions as the potential 
well of the star cluster gets shallower (cf. Figure \ref{fig:phitid}). In addition, a fraction of
particles is scattered from the SRLE region into the high-energy regions by two-body relaxation. 
The two-body relaxation leads to the small slope $0.127 \pm 0.030$ in Figure \ref{fig:baumgardt}.
It is small since $\alpha_M$ is very large for the models K1 - K9. From the values in 
Equation (\ref{eq:occupationnumbers}) we obtain $\alpha_M\approx 2$ for the model K9 with the
valid assumption that the particles of different mass are initially uniformly mixed in radius and 
Jacobi energy per unit mass.

If the physical tidal radius is equal to the radius where the 
density of the star cluster (King) model vanishes, we have $N_{\rm LRLE} = N_{\rm LRHE} = 0$
and only two of the four regions are occupied with particles. 
This is the standard case used in $N$-body simulations of star clusters
in a tidal field so far (e.g. Baumgardt \& Makino 2003, Trenti, Heggie \& 
Hut 2007, Ernst et al. 2007). In this case, $\alpha_M$ is small (typically a few percent)
and the dissolution times are considerably $N$-dependent as we checked with 
a few models ($N = 10^3, 2\times 10^3, 5\times 10^3, 10^4, 3\times 10^4$) using {\sc nbody6gc}. Furthermore, in this case our preliminary models 
suggest that the dissolution time directly scales with a power of the relaxation time. 
However, a more detailed study seems to be essential.
On the other hand, Tanikawa \& Fukushige (2005) adopted initial models
where the King cutoff radius was smaller or larger than the physical tidal radius.
By decreasing the size of the initial star cluster further $\alpha_M$ can be
forced to vanish.

We argue that the situation that the cluster is divided into the four regions 
of the RE classification 
(with certain occupation numbers and masses) is the typical situation for newly formed star clusters. 
A first crucial question is whether stars can form in all regions. 
The answer is yes, if the condition for star formation is fulfilled. According to the modern picture of 
gravo-turbulent star formation (e.g. Mac Low \& Klessen 2004,
Ballesteros-Paredes et al. 2007), 
supersonic turbulence and shocks create initial density enhancements in a molecular cloud. 
The formed molecular cloud core contracts gravitationally and fragments eventually. 
Finally, protostellar seeds form, accrete in-falling material and become main sequence stars. 
The condition for star formation is independent of the distinction between high- and 
low-energy regions of the effective potential. 
Thus one would expect that stars form initially in the high-energy regions and the SRLE 
region slowly builds up as more material moves towards the center of the new star cluster. 
Due to the turbulent structure within the molecular cloud it is also possible that a small fraction 
of stars forms in the LRLE region.

In the Galactic center, the supersonic shock and turbulent velocities must be high enough 
to form mean densities which withstand the tidal shear forces.
According to Morris (1993), the critical mean number density for 
gravitationally bound clouds in the Galactic center region is given by 

\be
n_{crit} = 10^7 \ \mathrm{cm}^{-3}  \left(\frac{1.6 \ \mathrm{pc}}{R_g}\right)^{1.8}, 
\ee

\noindent
where $R_g$ is the galactocentric radius. 

The picture sketched above would be similar if the star cluster formation in the Galactic center
is triggered by the collision of two clouds. For typical parameters (e.g. for the formation of
clusters like Arches and Quintuplet) the rate of such cloud collisions in the 
Galactic center is low as compared with the reciprocal of the lifetime of OB stars 
and can be crudely estimated to be 

\bea
\nu_{col} &=& 5\times 10^{-8} \ \mathrm{yr}^{-1} \left(\frac{M_{cloud}}{10^6 \ M_\odot}\right)^{-1} \nonumber \\ 
&&\times \left(\frac{N_{H_2}}{10^{23} \ \mathrm{cm}^{-2}}\right)^{-1} \left(\frac{\sigma_v}{20 \ \mathrm{km \ s}^{-1}}\right)
\eea

\noindent
where $M_{cloud}$, $N_{H_2}$ and $\sigma_v$ are the mass, the column density and the 
velocity dispersion of a cloud (Hasegawa et al. 1994, Stolte et al. 2008).

Finally, we note that the Jeans time scale is of the same order as the dissolution times
of our models in the Galactic center. According to  Hartmann (2002), 
who explored an earlier idea by Larson (1985), the Jeans 
(or fragmentation) time scale of a gaseous filament can be written as

\be
\tau \approx 3.7 \ \left(\frac{T}{10 \ \mathrm{K}}\right)^{1/2} A_V^{-1} \ \mathrm{Myr}
\ee

\noindent
where $T$ is the temperature and $A_V\approx 5$ is the visual extinction through the center 
of the filament (see also Klessen et al. 2004). 

\section{Discussion}

We have studied the dissolution of star clusters in an analytic background 
potential of the Galactic centre by means of direct $N$-body simulations. 
We described in detail the algorithm of our parallel $N$-body program {\sc nbody6gc}
which is based on Aarseth's series of $N$-body codes (Aarseth 1999, 2003, Spurzem 1999).
It includes a realistic dynamical friction force with a variable Coulomb logarithm based on the
paper by Just \& Pe\~narrubia (2005). The initial value for the circular orbit of the model K9 is 
$\ln\Lambda \approx 1.7$. It turns out that, even for a $10^6 \ M_\odot$ cluster, 
the dynamical friction force is too weak to let a cluster on a circular orbit at $R_C=20$ pc spiral 
into the Galactic centre within the lifetime of its most massive stars. Thus we did not resolve
the ``paradox of youth'' (Ghez et al.  2003).

However, we have studied in detail the dynamics of dissolving star clusters on circular
orbits in the Galactic centre. The key to the understanding of this dynamical 
problem is the gravitational potential which is the sum of the effective tidal
potential and the star cluster potential. Along the orbit of the star cluster, the effective
tidal potential resembles a parabolic wall. However, in the close vicinity of the Galactic 
centre there are deviations from the parabolic shape due to higher-order
terms in the Taylor expansion of the effective tidal potential. Due to this
asymmetry, the Lagrange points $L_1$ and $L_2$ lie at different energies.

We have studied in detail the properties of the tidal arms of a dissolving star cluster in a galactic
centre. The density wave phenomenon found by Capuzzo Dolzetta, di Matteo \& Miocchi (2005) 
and di Matteo, Capuzzo Dolcetta \& Miocchi (2005) appears in our model K9. 
The angles of the clumps can be calculated with the theory from Just et al. (2009).

We have presented a method to study the structure of tidal arms by using an eigensolver.
The eigensolver calculates numerically a 1D coordinate system along the
tidal arms and calculates characteristic dynamical quantities along this
coordinate system.

It may be of interest to note that more particles escape into the trailing tidal arm than into 
the leading tidal arm. For the high-energy particles the phase space for escape into
the trailing arm is larger than that for escape into the leading arm. 
The fractions of initial conditions in the phase space for escape into the leading and
trailing arm, respectively, have to be calculated numerically for several Jacobi energies.
The result would be called the `basins of escape' of the star cluster
(e.g. Aguirre et al. 2001, Contopoulos 2002, Ernst et al. 2008). This kind of asymmetry 
between the arms does not depend on the particle number.

The half-mass times of our models K1 - K9 depend only weakly on the particle number which 
indicates that two-body relaxation is not the dominant mechanism leading to the dissolution. 
The reason is that the initial models are divided into four different regions in radius and 
specific Jacobi energy space. 
This division has been termed the Radius-Energy (RE) classification. The division of a
newly formed star cluster into the four regions of the RE classification is probably a 
typical situation which is consistent with the modern picture of gravoturbulent
star formation (e.g. Mac Low \& Klessen 2004, 
Ballesteros-Paredes et al. 2007).
If the ratio $\alpha_M$ (which has been defined in Section \ref{sec:lifetimescaling}) is large 
enough, the dissolution is no longer relaxation-driven but the mass loss from the SRLE region
is governed by a self-regulating process of increasing Jacobi energy due to the weakening of 
the potential well of the star cluster, which is induced by the mass loss itself (Just et al. 2009). Predictions about the fractions of stars which belong to the four different regions
(i.e., the occupation numbers and masses) may be an important result of the emerging
theory of star cluster formation. What are typical ratios of occupation numbers and masses in
regions with efficient star formation? How do the occupation numbers, masses and
their ratios differ between open and globular clusters? 
From the side of stellar dynamics the scaling problem
of the dissolution times (Baumgardt 2001) needs to be
solved for the new dissolution mechanism due to a non-stationary gravitational
potential combined with the effect of two-body relaxation. In this paper, we also conjecture 
that the ratio $\alpha_M$ can be used to draw a distinction between associations, open and 
globular clusters, i.e. that the old globular clusters obey $\alpha_M \ll 1$ and that
their dissolution is relaxation-driven, while the open clusters and associations obey
$\alpha_M \gg 0$.


\section{Acknowledgements}

We thank Prof. Ortwin Gerhard and Dr. Kap-Soo Oh for many discussions related to
an earlier version of the program {\sc nbody6gc} and Dr. Godehard Sutmann for making us
aware of the composition schemes. Also, we thank Prof. Douglas Heggie and an anonymous 
referee for two comments which, together with a plot of Dr. Peter Berczik, finally led to the 
RE classification.

AE gratefully acknowledges  support by the International Max Planck Research School (IMPRS) 
for Astronomy and Cosmic Physics at the University of Heidelberg. 

We thank the DEISA Consortium ({\tt www.deisa.eu}), co-funded 
through EU FP6 projects RI-508830 and RI-031513, for support 
within the DEISA Extreme Computing Initiative.

\appendix

\section{Taylor expansion  of the effective tidal potential}

\label{sec:taylorexpansion}

\begin{figure}
\centering
\includegraphics[width=0.5\textwidth]{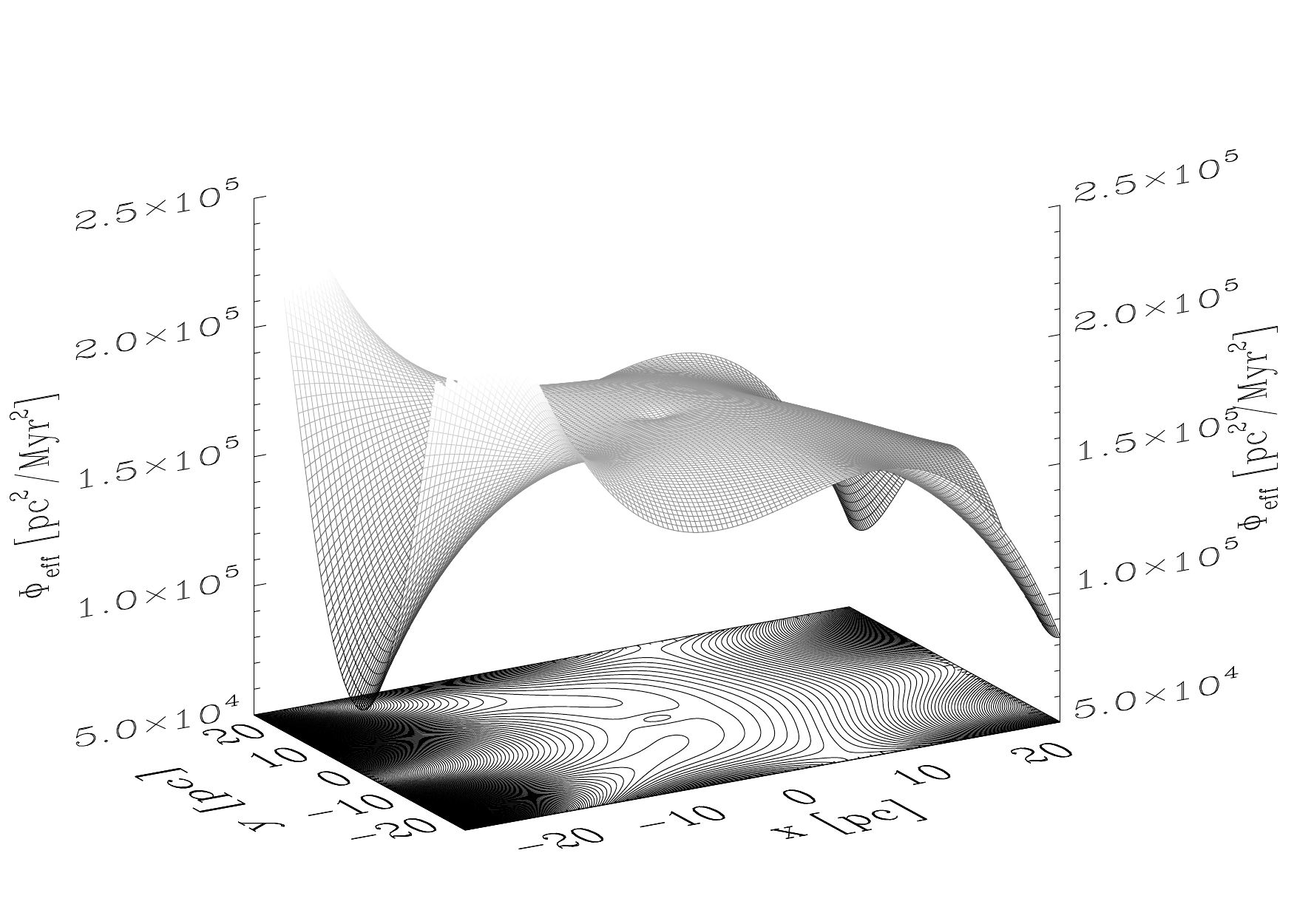} 
\caption{Taylor expansion of the effective tidal potential for 
the scale free model up to the $5$th order.}
\label{fig:teffpot}
\end{figure}

Usually a Cartesian Taylor expansion of the effective tidal potential is used.
Up to $5$th order, the 3D Cartesian Taylor expansion of the effective tidal potential for 
the scale free model is given by

\bea
\Phi_{\rm eff,tid} &\approx& \frac{1}{2} \left(\frac{3-\alpha}{\alpha -1}\right) \omega_0^2 R_0^2
+ \frac{1}{2} (\alpha - 3) \omega_0^2 x^2 \nonumber \\
&+& (\alpha-3)(\alpha-2)\frac{ \omega_0^2}{6 R_0} x^3 \nonumber \\
&+& (\alpha -3)\frac{\omega_0^2}{2 R_0} x (y^2 + z^2)  \nonumber \\
&+& (\alpha-4)(\alpha-3)(\alpha-2)\frac{\omega_0^2}{24 R_0^2} x^4 \nonumber \\
&+& (\alpha-4)(\alpha-3)\frac{\omega_0^2}{4 R_0^2} x^2 (y^2+z^2) \nonumber \\
&+& (\alpha-3)\frac{\omega_0^2}{4 R_0^2} y^2z^2 \nonumber \\
&+& (\alpha-5)(\alpha-4)(\alpha-3)(\alpha-2)\frac{\omega_0^2}{120 R_0^3} x^5 \nonumber \\
&+& (\alpha-5)(\alpha-4)(\alpha-3)\frac{\omega_0^2}{12 R_0^3} x^3 (y^2+z^2) \nonumber \\
&+& (\alpha-5)(\alpha-3)\frac{\omega_0^2}{8 R_0^3} x (y^4+z^4) \nonumber \\
&+&  (\alpha-5)(\alpha-3)\frac{\omega_0^2}{4 R_0^3} x y^2 z^2 
\eea

\noindent
where $R_0$ and $\omega_0$ are the radius and the frequency of the circular orbit.
This solution is shown in Figure \ref{fig:teffpot} (which may be compared with Figure
\ref{fig:effpotgc}). It can be seen that this Taylor expansion cannot be used to study the 
properties of tidal arms in the Galactic centre. For extended tidal tails cylindrical
coordinates should be used and for the radial asymmetry the exact potential.

The expansion up to the second order coincides with the tidal approximation. 
We have for the scale free model $(\alpha-3)\omega_0^2/2 = (\kappa_0^2 - 4\omega_0^2)/2$
which is the coefficient of the second-order term in the tidal approximation, where $\kappa_0$ is 
the epicyclic frequency.

\section{Tidal arm coordinate system}

\label{sec:tidalarmcoordinatesystem}

Based on routines from Numerical recipes (NR, Press et al. 2001), we developed the 
eigensolver {\sc eigentid} which calculates numerically a 1D coordinate 
system along the tidal arms and evaluates characteristic dynamical quantities along
this coordinate system. We denote the coordinate along the tidal arms as $w$, where negative
values refer to the leading arm and positive values to the trailing arm. The NR routine {\tt tred2} uses the Householder reduction of a real symmetric $n\times n$ matrix to convert it to a tridiagonal form. The NR routine {\tt tqli} uses the QL algorithm to determine
the eigenvalues and eigenvectors of the matrix which has been brought into tridiagonal form before
(see NR, Chapters 11.2 and 11.3). 
We use the tensor of inertia and denote the eigenvectors corresponding to the minimum eigenvalue $c$, the medium eigenvalue $b$ and the maximum eigenvalue $a$ as the minimum, medium and maximum eigenvectors,
respectively. Then the algorithm proceeds as follows:

\begin{enumerate}
\item Read snapshot with particle masses, positions and velocities in the cluster rest frame.
\item Calculate optionally gravitational potential and density (using the method by Casertano \& Hut 1985)
for this snapshot.
\item Start calculation in the origin of coordinates $(0,0,0)$.
\item Obtain a neighbour sphere with radius $R_{\rm cut}$ and calculate its centre of mass $(x_{cm}, y_{cm}, z_{cm})$.
\item Calculate physical quantities averaged over the neighbour sphere: 
Mean specific angular momentum, mean specific energy, mean density,
velocity dispersion, mean potential. Write all quantities to a data file.
\item Calculate the tensor of inertia of the neighbour sphere with respect to the centre of mass
of the neighbour sphere. It is given by

\begin{equation}
\Theta_{jk} = \sum_{i=1}^{N_{nb}} m_i \left( \begin{array}{lll}
\Delta y_i^2 + \Delta z_i^2 & \Delta x_i \Delta y_i & \Delta x_i \Delta z_i \\
\Delta x_i \Delta y_i & \Delta x_i^2 + \Delta z_i^2 &  \Delta y_i \Delta z_i \\
\Delta x_i \Delta z_i &  \Delta y_i \Delta z_i &   \Delta x_i^2 + \Delta y_i^2 \\
\end{array}\right)\label{eq:inertiatensor}
\end{equation}

\noindent
where $\Delta x_i = x_i-x_{cm}, \Delta y_i = y_i - y_{cm}$ and $\Delta z_i = z_i - z_{cm}$ are the relative positions of the $i$th particle in the neighbour sphere with respect to its centre of  mass and
$m_i$ is the mass of the $i$th particle.
\item Calculate the eigenvalues and eigenvectors of $\Theta_{jk}$.
\item Go along the direction of the maximum/medium eigenvector until a critical density is reached to find the new  $R_{\rm cut}$.
\item Check for acute angle between previous and current minimum eigenvector. If the angle is acute, change the sign of the eigenvector.
\item Go one step along the direction of the minimum eigenvector.
\item Repeat from 4. until the particle number within the neighbour sphere drops below a certain threshold as the first tidal arm ends.
\item Start from 3. for the second tidal arm.
\end{enumerate}

\noindent
We remark that the inertia ellipsoids of the neighbour spheres have an oblate shape, i.e.
the three eigenvalues $a,b,c$ of $\Theta_{jk}$ satisfy $a\approx b > c$. Also, a weighting
exponent can be assigned to the particle mass in the expression (\ref{eq:inertiatensor}).
In this case, the eigensolver follows the mass distribution within the tidal arms in a 
different way. This method has been applied for Figure \ref{fig:disperse100kb}.

\bsp

\label{lastpage}

\end{document}